\documentclass[%
prx,
english,
twocolumn,
amssymb,
aps,
longbibliography,
groupaddress, nofootinbib]{revtex4-2}

\usepackage{graphicx}
\usepackage{natbib}
\usepackage{amsmath}
\usepackage{amssymb}
\usepackage{amsfonts}
\usepackage{mathrsfs} 
\usepackage{dsfont} 
\usepackage{xcolor}
\definecolor{darkblue}{rgb}{0,0,.65}
\definecolor{darkgreen}{rgb}{0.28,0.41,0.19}
\definecolor{nicegreen}{rgb}{0.28,0.85,0.19}
\usepackage{lipsum}
\usepackage{nicefrac}
\usepackage{bm} 
\usepackage{enumitem}
\usepackage{physics}

\usepackage[%
  pdfauthor={Benedikt Placke},%
  pdfstartview=FitH,%
  breaklinks=true,%
  bookmarks=true,
  colorlinks=true,%
  anchorcolor=black,%
  citecolor=darkgreen,
  filecolor=black,%
  menucolor=black,%
  urlcolor=darkblue,%
  linkcolor=blue,%
 ]{hyperref}

\def\equationautorefname~#1\null{Eq. (#1)\null}

\newcommand{\appref}[1]{\hyperref[#1]{App.~\ref*{#1}}}

\renewcommand\vec{\bm}

\DeclareMathOperator{\pauligroup}{\mathcal P_N}
\DeclareMathOperator{\id}{\mathds{1}}
\DeclareMathOperator{\reals}{\mathbb{R}}
\DeclareMathOperator{\linspan}{\mathrm{Span}}
\newcommand{\vecspace}{\mathscr}

\begin{document}

\title{Slow measurement-only dynamics of entanglement in Pauli subsystem codes}

\author{Benedikt Placke}\email{benedikt.placke@physics.ox.ac.uk}

\author{S. A. Parameswaran}\email{sid.parameswaran@physics.ox.ac.uk}

\affiliation{Rudolf Peierls Centre 
for Theoretical Physics, 
University of Oxford, Oxford OX1 3PU, United Kingdom}

\date{\today}

\begin{abstract}
We study the non-unitary dynamics of a class of quantum circuits based on stochastically measuring check operators of subsystem quantum error-correcting codes, such as the Bacon-Shor code and its various generalizations. Our focus is on how properties of the underlying code are imprinted onto the measurement-only dynamics. 
We find that in a large class of codes with nonlocal stabilizer generators, at late times there is generically a nonlocal contribution to the subsystem entanglement entropy which \emph{scales} with the subsystem size. 
The nonlocal stabilizer generators can also induce slow dynamics, since depending on the rate of competing measurements the associated degrees of freedom can take exponentially long  (in system size) to purify (disentangle from the environment when starting from a mixed state) \emph{and} to scramble (become entangled with the rest of the system when starting from a product state).
Concretely, we consider circuits for which the nonlocal stabilizer generators of the underlying subsystem code take the form of subsystem symmetries. We present a systematic study of the phase diagrams and relevant time scales in two and three spatial dimensions for both Calderbank-Shor-Steane (CSS) and non-CSS codes, focusing in particular on the link between slow measurement-only dynamics and the geometry of the subsystem symmetry. A key finding of our work is that slowly purifying or scrambling degrees of freedom appear to emerge only in codes whose subsystem symmetries are nonlocally {\it generated}, a strict subset of those whose symmetries are simply nonlocal. We comment on the link between our results on subsystem  codes and the phenomenon of Hilbert-space fragmentation in light of their shared algebraic structure.
\end{abstract} 

\maketitle


\section{Introduction\label{sec:intro}}

Quantum error correction (QEC) is essential in order to enable the robust processing of quantum information. It played a key role in establishing that scalable fault-tolerant quantum computation is possible in principle, and remains of central {practical} importance in the  quest to move beyond the ``noisy, intermediate-scale quantum'' (NISQ) regime~\cite{Preskill2018NISQ} achieved by present-day platforms. Despite these pragmatic origins, QEC is deeply linked to many fundamental problems in quantum many-body physics. For example, the concept of topological order in equilibrium quantum systems is intimately related to the non-recoverability of local information in a quantum code~\cite{kitaev_tc,beyond}, while the search for a self-correcting quantum memory led to the identification of various types of fracton topological order~\cite{yoshida_no_strings,haah_code,vijay_fto}, with implications both for equilibrium phase structure and glassy dynamics out of equilibrium.
Error correction itself is  a concrete instance of what has sometimes been dubbed ``quantum interactive dynamics'' \cite{Sierant2022dissipativefloquet,wu2023qid,mcginley2022qca,machado2023qca,friedman2023mipt_adaptive,odea2024qid,ravindranath2023qid}, where the future time evolution of a system is influenced by the knowledge acquired, via measurements, about its present state: in this setting, quantum  error correcting codes can be viewed as undergoing dynamical phase transitions as the error rate is varied~\cite{dklp,bao2023mixedstate,deGroot2022symmetryprotected,wang2024intrinsic,sohal2024noisy,placke2023hyperbolic}.

Recent attention has focused on the \emph{dynamical generation} of quantum error correcting codes. This builds on the idea that characterizing the (late-time) state of a system can be fruitfully viewed as a problem in  error correction. 
This task is particularly simple when the local dynamics take a restricted ``Clifford'' form, since in this case at any time the dynamical state of a system of $N$ qubits initialized in an appropriately chosen computational basis state is the simultaneous eigenstate of $N$ commuting operators from the Pauli group $\pauligroup$ (though the set of operators can itself change). Any such ``stabilizer group'' naturally defines a quantum code, whose properties can be used to characterise the state itself.
This paves the way to generating previously unknown codes through dynamics. These can be produced either by unitary circuits built from local few-qubit gates arranged in a random~\cite{brown2021scrambling_speed}  or deterministic~\cite{sommers2023crystalline} pattern in space-time, or through sequences of projective measurements. In the latter case, again, operators can be measured either randomly in space and time~\cite{ippoliti2021MOD}, or specifically-chosen ones can be measured according to a fixed time-periodic schedule  --- as in the celebrated ``Floquet codes'' ~\cite{hastings_haah_floquet, floquet_anyon_condensation_a, floquet_anyon_condensation_b}.
This framework also offers a useful alternative perspective on circuits which include both unitary gates and measurements. For example, the measurement-induced phase transition observed originally in such a hybrid unitary-projective setting~\cite{li2018mipt,li2019MIPT,skinner2019MIPT} can be understood as an error-correction transition where information scrambling by unitary evolution in a dynamically generated code either succeeds or fails in concealing information from projective measurements, leading to differing scaling behaviour of entanglement in the time-evolved state \cite{choi2020mipt_qec}.

\begin{figure*}
\centering{}
\includegraphics{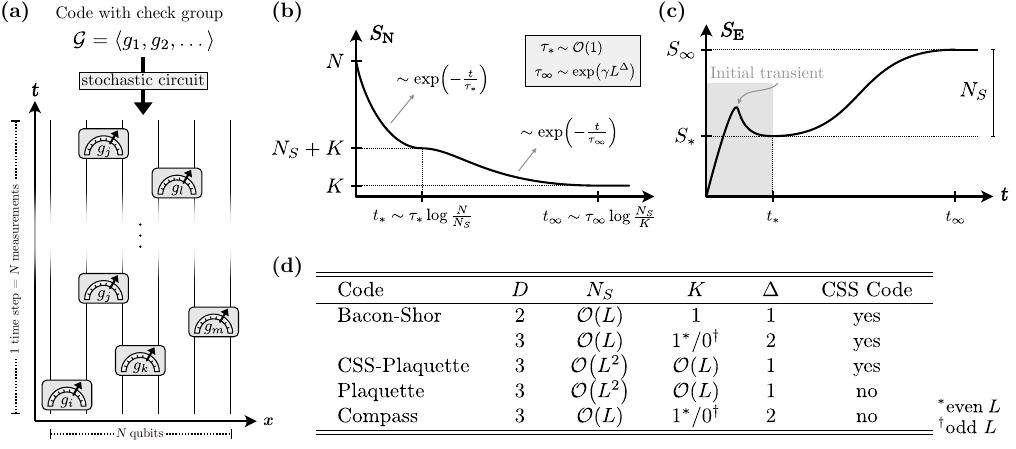}
\caption{
(a) Measurement-only circuits based on subsystem codes. A subsystem code is defined by a check group $\mathcal G$, which is generated by geometrically local check operators $g_j$ acting on $N$ qubits. In the corresponding circuit, we measure $N$ randomly chosen check operators per time step. 
We sketch the purification and scrambling dynamics of such circuits in panels (b) and (c), respectively. 
We observe a separation of typical time scales for both the decay of the von Neumann entropy $S_N$ and for the growth of the half-system entanglement entropy $S_{\rm E}$. The larger time scale $t_{\infty}$, which is exponentially large in a subextensive power of the linear system size, arises if the stabilizer group (see \autoref{fig:algebras}) of the underlying subsystem code has $N_{S} > 0$ non-local generators (which may exist even if all measured operators are geometrically local). 
In panel (d), we tabulate all codes analysed in this work.
All the codes we study are defined on a hypercubic lattice of dimension $D$ (so $N=L^D$) and the $N_S$ nonlocal stabilizer generators take the form of subsystem symmetries acting on subsystems of dimension $\Delta$ (so $N_S\sim L^{D-\Delta}$). The number of logical qubits is denoted by $K$.
}
\label{fig:intro}
\end{figure*}

Here, we draw inspiration from these developments but study the converse question: rather than generating codes {\it from} dynamics, we instead  study the {\it dynamics generated by codes}. In other words, we build a specified code structure into  a measurement-only quantum circuit, and examine how this structure is imprinted onto the properties of its steady state and late-time dynamics. 

To do so, we consider codes defined by a specified set of ``check operators'' and measure a randomly chosen check operator at each step in time.  For the best-known class of error-correcting codes --- stabilizer codes, all of whose check operators commute --- this results in trivial dynamics. In order to generate richer structure, measurements of stabilizers can be interspersed with those of single qubits, leading to complex phase diagrams~\cite{lavasani2021topological_1D,SangHsiehMeasurementProtection,lavasani2021topological_2D}; the link to error correction is then via the resilience of the original code to single-qubit noise~\cite{botzung2023robustness,lee2024randomly_monitored_codes}.
This is a rich subject in its own right; however, the addition of single-qubit measurements necessarily breaks the algebraic (symmetry) structure inherited by the circuit from its underlying quantum code.
In order to generate nontrivial dynamics while preserving code structure, we instead study measurement-only circuits based on \emph{subsystem codes}. Such codes are defined in terms of a group of check operators $\mathcal G$ which do not, in general,  commute. As is illustrated in \autoref{fig:intro} (a), choosing a generating set $\{g_j\}$ for $\mathcal G$, at each time step we randomly measure one of these generators (for convenience we choose a unit of time in which $N$  measurements are possible). Since the $g_j$ do not generally commute, such circuits can have nontrivial measurement-only dynamics in their own right but in addition they inherit from the code nontrivial symmetries in the form of its stabilizer and logical operators (which together form the centralizer $\mathcal C_{\pauligroup}(\mathcal G)$ of $\mathcal G$). 
These symmetries strongly constrain many aspects of the dynamics; understanding precisely how is our central focus here.

Although individual examples of this kind have been studied previously~\cite{botzung2023robustness,kitaevMO2023a,kitaevMO2023b}, a comprehensive analysis of how spatial dimension and stabilizer geometry --- among other properties of the underlying codes --- manifest themselves in the approach towards and time evolution within the steady state of such dynamics has been lacking. It is this overarching understanding that we supply in the present work.
More generally, our work complements earlier studies on the role of on-site symmetries in monitored circuits \cite{barratt2022charge_sharpening, agrawal2022charge_sharpening, majidy2023spin_sharpening}.

Our main findings are that for a large class of subsystem codes, the resulting measurement-only circuit dynamics (i) exhibit nontrivial steady states whose bipartite entanglement entropy has \emph{leading-order} nonlocal contributions, even in cases where it satisfies area-law scaling with subsystem size; (ii) can show very slow approaches to steady-state scaling of the entanglement entropy, only saturating it at a time scale that can be typically exponentially large in (a subextensive power of) the system size.
More precisely, both these features hold if the stabilizer group  of the subsystem code (the center of $\mathcal G$) underlying the circuit has \emph{nonlocal generators}. 
This results in two distinct dynamical regimes, perhaps most clearly visible in `purification dynamics': initializing the system in a maximally mixed initial state $\rho(t=0) =0$, and  tracking the von Neumann entropy $S_{\rm N}$ as a function of time. 
In this case, one can view the nonlocal stabilizer generators as encoding information about the state $\rho$ that can be learned by local check measurements (since stabilizers are, by definition, products of checks), but for which the typical time on which is learned by the dynamics can be long, depending on the rate of competing (i.e., non-commuting) measurements.
As sketched in \autoref{fig:intro} (b), the von Neumann entropy initially decreases exponentially with rate $\tau^{-1}_*$ independent of system size and plateaus, after a typical time of $t_* \sim \tau_* \log(N / (N_S+K))$, at a value of $N_S + K$, where $N_S$ is the number of nonlocal stabilizer generators that take long to be measured effectively and $K$ is the number of logical qubits of the code. This fast initial drop in entropy corresponds to the dynamics learning all information that is efficiently obtained by local measurements. 
After time $t_*$, the entropy continues decreasing as all nonlocal stabilizer generators are measured, but with a rate $\tau_{\infty}^{-1}$ that is exponentially small in system size. 
The infinite-time value, given by $S_{\rm N}(t\to\infty ) = K$ (since the logical degrees of freedom cannot purify) is then reached after a typical time $t_\infty = \tau_{\infty}\log(1+N_S / K)$.
These two distinct time scales are also probed by the subsystem entanglement entropy $S_{\rm E}$, as sketched in \autoref{fig:intro} (c). When starting an initial random product state, the system, after an initial transient, scrambles its locally available information at  time $t_*$, but picks up an additional contribution scaling with $N_S$ at exponentially long times $t_{\infty}$. This contribution is generally subleading but can contribute at leading order in area law phases.

We demonstrate the above picture in detail by giving physically intuitive arguments in the stabilizer formalism and by deploying extensive numerical simulations on an array of examples summarized in \autoref{fig:intro} (d). In all cases, the codes are defined on a hypercubic lattice of dimension $D$ (i.e. $N \sim L^D$), and \emph{all} stabilizer generators are nonlocal and take the form of subsystem symmetries acting on subsystems of dimension $\Delta$, with $N_S\sim L^{D-\Delta}$. 
We also study the phase diagrams of these codes as a function of relative measurement rates of different check operators and characterise how both the topology of the phase diagram as well as the scaling of $\tau_\infty$ in these phases depends on the geometry of the subsystem symmetry. 
We find that while in circuits with line-like stabilisers ($\Delta = 1$) there exists a single dynamical transitions that separates two distinct area-law phases where different subsets of the full stabilizer group purify slowly, in circuits with sheet-like stabilizers ($\Delta=2$) there may exist an intermediate phase where \emph{all} stabilizer generators purify slowly. In both cases, the time scale on which slow stabilizers are measured scales as $\tau_\infty\sim\exp(\gamma L^\Delta)$. 
We also compare circuits based on CSS and non-CSS codes and find that only the latter generate steady states with a volume-law scaling of the entanglement entropy as part of their phase diagram, in consonance with a recent conjecture \cite{ippoliti2021MOD}.
Intriguingly, our work also reveals a connection between measurement-only subsystem code dynamics and the phenomenon of Hilbert space fragmentation observed for \emph{unitary} dynamics, and hence our results may be regarded as the first step towards a more general understanding of how symmetries can also have non-trivial consequences for non-unitary quantum dynamics.

The rest of the paper proceeds as follows. In \autoref{sec:mod-and-ssc} we introduce subsystem quantum error correction codes, explain how we construct corresponding measurement-only circuits, derive the relation between code properties and circuit dynamics and discuss the relation to Hilbert space fragmentation. We begin \autoref{sec:results} by giving an overview of our results. We then present several  concrete examples in two and three spacial dimensions that help to both motivate and establish these results. We consider circuits with string- and membrane-like subsystem symmetries, as well as examples based on CSS and non-CSS codes, and explain the consequences of these choices on the phase diagrams of our models, and on purification and scrambling dynamics. We conclude in \autoref{sec:conclusion} with a summary and a discussion of possible future directions.

\section{Background and methods\label{sec:mod-and-ssc}}

In this paper, we focus on stochastic measurement-only dynamics~\cite{ippoliti2021MOD}: a discrete time evolution where at each time step a (geometrically) local operator is drawn from a specified distribution and projectively measured. Throughout, we will consider systems of $N$ qubits, and will choose the operators uniformly from a set of ``check'' operators 
of a given subsystem code (these terms are defined below). Unless specified otherwise, we fix the units of time $t$ such that $N$ operators have a chance to be measured in an interval $\Delta t=1$. This setup is also sketched in \autoref{fig:intro} (a).

In the remainder of this section, we first introduce the notion of subsystem codes and discuss their algebraic structure and  its link to  
the commutant-algebra formulation of the phenomenon of Hilbert-Space fragmentation in {\it unitary} dynamics~\cite{moudgaya2022fragmentation}. We then introduce the measurement-only dynamics and entanglement measures deployed in the rest of this paper. Readers familiar with these concepts could either omit this section or simply skim it in order to familiarize themselves with our notation, and skip ahead to our results in Section~\autoref{sec:results}

\subsection{Stabilizer Subsystem Codes: Algebraic Structure}

\begin{figure}
\includegraphics{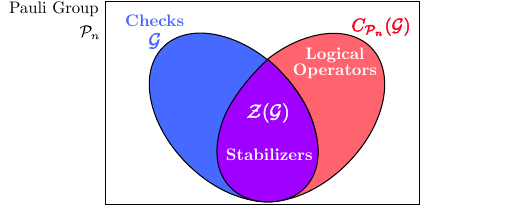}
\caption{Algebraic structure of stabilizer subsystem codes. The check operators generate the group $\mathcal G$. Its center, $\mathcal Z(\mathcal G)$, is the stabilizer group of the code. The bare logical operators of the code are given by $\mathcal C_{\pauligroup}(\mathcal G) - \mathcal Z(\mathcal G)$, where $C_{\pauligroup}(\mathcal G)$ is the centralizer of $\mathcal G$ in the $N$-qubit Pauli group $\pauligroup$. }
\label{fig:algebras}
\end{figure}

Subsystem codes are generalizations of the perhaps better-known  {\it stabilizer codes}, whose properties we first summarize to serve as a point of comparison.
A stabilizer code is defined by fixing an abelian subgroup of the Pauli group $\pauligroup$ on $N$ qubits, termed the stabilizer group $\mathcal S$ of the code. This is usually done by specifying a set of mutually commuting stabilizer generators $S_j$ such that $\mathcal S = \expval{S_j, \, j=1\dots N_S}$. 
This induces a decomposition of the full Hilbert space of the form 
\begin{equation}
    \vecspace{H} = \bigoplus_{\vec \sigma \in \mathbb Z_2^{N_S}} \vecspace{C}(\vec \sigma), \label{eq:stabdecomp}
\end{equation}
where the sum is over all $2^{N_S}$ possible bitstrings $\vec\sigma$  of eigenvalues of the stabilizer generators $S_j$. (In other words, the Hilbert space decomposes into a direct sum of subspaces with fixed values of the $N_S$ stabilizers.) Logical information is encoded into one of these subspaces, termed the `code space'. This is conventionally taken to be the $+1$ eigenspace of all the stabilizers, although practical error correction routines typically infer the set of stabilizer eigenvalues (not necessarily all $+1$) and performs appropriate classical post-processing whenever possible. 

Stabilizers by definition act trivially on the code space and hence represent the logical identity. Operators which preserve the code space but act nontrivially on it are called logical operators. For stabilizer codes, logical operators are exactly those operators that commute with all stabilizers but cannot themselves be written as a product of stabilizers. In other words, the group of logical Pauli operators is the centralizer\footnote{Recall that the centralizer or commutant of a subset $\mathcal S$ of a group $G$ is the set of $C_G(\mathcal{S})$ of elements in $G$ that commute with every element in $S$; equivalently, $gsg^{-1} =s$ for every $s\in\mathcal{S}$ and $g\in C_G(\mathcal S)$.} of $\mathcal S$ in $\pauligroup$,  denoted $C_{\pauligroup}(\mathcal S)$.  An automatic corollary is that multiplication by stabilizers induces an equivalence relation between logicals: two logicals that only differ by a product of stabilizers are identified. When linked with an appropriate notion of locality (e.g., on a lattice), this equivalence relation endows the code with an appealing topological interpretation, as in the case of the toric or surface codes or even more exotic fracton models.   
 The dimension of the logical subspace is $\rm dim \vecspace C = 2^K$, where $K$ is the number of logical qubits, and the support of the smallest nontrivial logical operator [i.e., element of $C_{\pauligroup}(\mathcal S)$] is called the code distance. A stabilizer code with $N$ physical qubits, $K$ logical qubits, and distance $D$ is denoted an $[[N, K, D]]$ code. 

In an extension of this  structure, subsystem codes~\cite{kribs2006sscodes} encode logicals not in a subspace of the total Hilbert space, but instead in a logical subsystem (hence the name). Focusing specifically on the case of stabilizer\footnote{More general subsystem codes only demand the decomposition of the Hilbert space as $\vecspace H = \vecspace R \oplus (\vecspace G \otimes \vecspace L)$.} subsystem codes, this corresponds to a further decomposition of the Hilbert space beyond the direct-sum structure of a stabilizer code, viz.

\begin{equation}
 \vecspace H = \bigoplus_{\vec \sigma \in \mathbb Z_2^{N_S}} \vecspace{L}(\vec \sigma) \otimes \vecspace{G}(\vec \sigma),
    \label{eq:h-ssc}
\end{equation}
where $\vecspace L$ and $\vecspace G$ are respectively termed  the logical and  gauge  subsystems (The latter terminology is historical and should not be confused   with the `gauging'  that arises when codes are studied in physical settings; sometimes $\vecspace G$ is termed the `junk' or `garbage' subsystem for this reason.)  
The decomposition in \autoref{eq:h-ssc} is induced by specifying a set of check operators $g_j \in \pauligroup$ which, in contrast to stabilizer generators, need not mutually commute. 
The checks generate the gauge group $\mathcal G$, and the center $\mathcal Z (\mathcal G)$ of the gauge group ----- i.e., its maximal abelian subgroup --- is identified with the stabilizer group of the subsystem code. The subspaces $\vecspace{C}(\vec\sigma)= \vecspace L(\vec\sigma) \otimes \vecspace G(\vec\sigma)$ are labeled the eigenvalues $\vec\sigma$ of all stabilizer generators. Conventionally, the subspace used for the code is the one with $\vec \sigma = (1\dots 1)$, that is the common +1 eigenspace of all stabilizers. Evidently, when $\mathcal{G}$ is abelian so that $\mathcal{G} = \mathcal{Z}(\mathcal{G})$, we recover the stabilizer group structure of \autoref{eq:stabdecomp}. We can thus view a stabilizer subsystem code as arising from decomposing  each stabilizer subspace of a stabilizer code into gauge and logical subsystems. 

The logical operators of the subsystem code  are those operators that leave each logical subsystem $\vecspace L(\vec\sigma)$ invariant. However, in contrast to stabilizer codes, logical operators of a subsystem code are equivalent up to multiplication with gauge operators in $\mathcal G$, rather than just up to multiplication with stabilizers. A particular set of representatives of the logical operators of the code are the so-called bare logical operators. These act trivially on the gauge subsystem $\vecspace G$, that is for any bare logical $L$, we can write 
\begin{subequations}
\begin{equation}
L = \bigoplus_{\vec \sigma \in \mathbb Z_2^{N_S}} L_{\vecspace L(\vec \sigma)} \otimes I_{\vecspace G (\vec \sigma)}
\end{equation}
 where $I_{\vecspace V}$ is the identity operator on vector space $\vecspace{V}$. 
Stabilizers are trivial bare logical operators, while in general the bare logical operators are the elements of the centralizer of the gauge group $C_{\pauligroup}(\vecspace G)$. Any logical operator can then be written as  the product of a bare logical operator and a gauge operator.
For completeness, we mention that by definition the gauge group preserves the logical subspace so 
\begin{equation}
g = \bigoplus_{\vec \sigma \in \mathbb Z_2^{N_S}} I_{\vecspace L (\vec \sigma)} \otimes g_{\vecspace G(\vec \sigma)}
\end{equation}
\label{eq:op-ssc}
for any check operator $g\in \mathcal G$.\footnote{In fact, as will become important in the next section, the decompositions in \autoref{eq:op-ssc} are possible not only for the elements of $C_{\pauligroup}(\vecspace G)$ and $\mathcal G$, but also for elements of their respective linear spans.}
The algebraic (operator) structure of subsystem codes is sketched in \autoref{fig:algebras}.
\end{subequations}

As before, $\dim \vecspace C = 2^K$, where $K$ is the number of logical qubits of the code and its distance is the support of the smallest nontrivial logical operator. Importantly, for subsystem codes this includes dressed (that is non-bare) logicals. The dimension of the gauge subspace is usually written as $\dim [\vecspace G(\vec\sigma)] = \dim (\vecspace G) = 2^R$, where $R = N - K - N_S$ is the number of gauge qubits and $N_S$ as before denotes the number of independent stabilizer generators, while that of the logical subspace is given by $\dim[\vecspace L(\vec \sigma)] = \dim(\vecspace L)=2^K$. Note that the dimensions of both the gauge and logical subspaces are independent of the stabilizer eigenvalues.
 A subsystem code with $N$ physics qubits, $K$ logical qubits, $R$ gauge qubits, and distance $D$ is called an $[[N, K, R, D]]$ code. 

A Calderbank-Shor-Steane (CSS) code~\cite{calberbank1996qec,steane1996qec} is a special case where all the check operators $g_j$ of the code are either products of only Pauli-$X$ or Pauli-$Z$ operators. A well-known example of a CSS subsystem code is the Bacon-Shor code\cite{bacon2006bsc}, while a well-known example of a non-CSS subsystem code is provided by Kitaev honeycomb model; however when viewed as a subsystem code the latter has a trivial (i.e. zero-dimensional) logical subspace~\cite{Suchara2011kitaev_ssc}.

\subsubsection{Connection to Hibert-Space Fragmentation}

We also comment  on a link between the algebraic structure of subsystem codes  and the bond-algebraic perspective on the phenomenon of Hilbert space fragmentation~\cite{moudgaya2022fragmentation}. To make the connection precise, consider a Hamiltonian $H = \sum_j \alpha_j g_j$, where $\alpha \in \reals$ and the $g_j\in \mathcal G$ are the check operators of a given stabilizer subsystem code.
The time evolution operator\footnote{Note that a similar structure can also be defined for general unitary time evolution operators that lack a well-defined local Hamiltonian, e.g. in a Floquet system, but we discuss the Hamiltonian version for concreteness.} $U(t) = \exp(-i t H)$ is then in the linear span $\linspan\mathcal G$ of elements in the gauge group, which for Hamiltonian systems is termed the bond algebra, and hence preserves both the full set of stabilizer eigenvalues as well as the state of the logical subsystem of the code. 

\begin{figure}
    \includegraphics{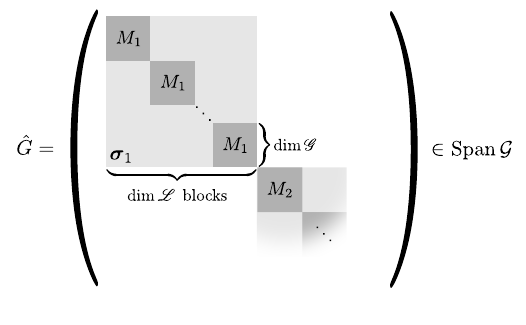}
    \caption{Block-diagonal structure of the matrix representation of an element $\hat G$ of the gauge algebra $\linspan{\mathcal G}$ in an appropriately chosen tensor-product basis $\ket{\vec \sigma, \vec z_b} \otimes \ket{\vec \sigma, \vec z_l}$ [see main text 
    for details]}
    \label{fig:ssc-blocks}
\end{figure}

To draw the parallel in full, first note that the decompositions shown in \autoref{eq:op-ssc} are in fact not only possible for the bare logical operators $C_{\pauligroup}(\vecspace G)$ and the check operators $\mathcal G$, but also for any element in their respective linear span $\linspan C_{\pauligroup}(\vecspace G)$ and $\linspan \mathcal G$. 
Returning to the Hamiltonian example, this means in particular that any element of the bond algebra $\linspan\mathcal G$, and hence also $U(t)$, is block diagonal as sketched in \autoref{fig:ssc-blocks}, with $2^{N-R}$ blocks of dimension $2^R$ when written in an appropriate product basis $\ket{\vec \sigma, \vec z_b} \otimes \ket{\vec \sigma, \vec z_l}$. The bitstrings $\vec z_b$ and $\vec z_l$ respectively specify bases of the gauge and logical subsystems for a given stabilizer sector labelled by $\vec \sigma$.  
For the case of the time evolution operator $U(t)$, the block-diagonal structure implies the existence of $2^{N-R}$ invariant subspaces, which are called Krylov subspaces in the context of Hilbert space fragmentation~\cite{pai2019fracton_circuits,sala2020fragmentation, khemani2020fragmentation}. From the preceding discusson, we see that this corresponds precisely to the decomposition in \autoref{eq:h-ssc}: in other words, the Hilbert space of a subsystem code is `fragmented' according to this bond-algebraic definition~\cite{moudgaya2022fragmentation}. [We observe that the fact that code Hamiltonians can naturally  \emph{evade} of thermalization provides an elegant counterpoint to the fact that eigenstates of \emph{chaotic} Hamiltonians obeying the Eigenstate Thermalization Hypothesis (ETH) form approximate quantum error-correcting codes
\cite{*[{To the best of our knowledge, this observation first appeared in print in a clear form in }] [{. However, S.A.P. recalls being educated on the parallels between ETH and the Knill-Laflamme conditions for QEC by Steve Flammia and the late David Poulin during an illuminating 2015 discussion in Moorea, French Polynesia.}] brandao2019qec_eth}.]

The fact that such a structure in the  bond algebra of a family of Hamiltonians naturally leads to the emergence of multiple Krylov subspaces even within a single global symmetry sector, and the consequences for unitary dynamics (which include the possibility of ergodicity breaking)  was first noted in Ref.~\cite{moudgaya2022fragmentation} (which also mentioned but did not systematically explore the connection to subsystem codes). This is believed to underpin various exotic dynamical phenomena in a range of isolated quantum systems, and has been experimentally explored  in the setting of one-dimensional Rydberg arrays.  Below, we show that with an additional condition on the stabilizer geometry, one can define a measurement-only analogue of the `anomalous' dynamics often associated with Hilbert-space fragmentation in the unitary setting.

\subsection{Stochastic measurement-only dynamics\label{sec:stochmeasdyn}}

We now define a class of stochastic measurement-only dynamics~\cite{ippoliti2021MOD} that can be associated with a specified subsystem code. In each time step, a random check operator of the subsystem code is measured. Different check operators will in general be measured with different rates, while preserving translational invariance of the ensemble. 
Such dynamics has been considered previously, in particular in Ref.~\onlinecite{sharma2023BaconShor}, which studies measurement-only circuits based on the Bacon-Shor code (see \autoref{sec:2d_bsc} for a definition) and uncovers an interesting spin-glass like structure of correlation functions in the ensemble of steady states. However  Ref.~\onlinecite{sharma2023BaconShor} did not explore the purification dynamics and slow entanglement growth that will be the main topics of this paper. Recent studies of the measurement-only Kitaev model~\cite{kitaevMO2023a,kitaevMO2023b,zhu2023kitaev_color_code} also highlight its interpretation as a subsystem code. This is  useful for gaining intuition about the dynamics, despite the fact that the Kitaev model (when viewed as a subsystem code) does not encode a logical qubit and is hence trivial.

We focus specifically on Pauli stabilizer 
subsystem codes since their dynamics can be efficiently  computed numerically  using the stabilizer formalism~\cite{gottesmann1998stabilizer, aaronson2004stabilizer},
by exploiting the fact  that evolution under a restricted set of operations can be efficiently computed in the Heisenberg picture, as we now describe. First, consider a state described by the density matrix $\rho$ that is the $+1$ eigenstate of a set of independent and mutually commuting Pauli operators $s_i\in\pauligroup$, $i = 1\dots m$. The operators $s_i$ form the minimal generating set of the stabilizer group $\mathcal S(\rho)$ of the state $\rho$. A pure state on $N$ qubits is uniquely fixed by a stabilizer group with $N$ independent generators. More generally, if the stabilizer group $\mathcal S(\rho)$ has $m < N$ generators, the state is the maximally mixed state on the common $+1$ eigenspace of $\mathcal S(\rho)$, so that
one can identify
\begin{equation}
    \rho = \frac{1}{2^N} \sum_{s \in \mathcal S(\rho)} s.
    \label{eq:stab-rho}
\end{equation}
The von Neumann entropy of the state is then $S = - \tr\rho\log_2\rho = N - m$. Evidently, we can specify a state by simply listing a (non-unique) set of generators $\{s_1,s_2 \ldots,s_m\}$ for its stabilizer group, and $\mathcal{S}(\rho) =\expval{s_1, s_2, \ldots, s_m}$, where $\expval{\ldots}$ denotes the group generated by the listed stabilizers. [Note that this introduces a second notion of stabilizer group, namely that of  the stabilizer group of a state, $\mathcal S(\rho)$, which is an efficient description of its density matrix in the sense of \autoref{eq:stab-rho}. This is distinct from the previously-encountered notion of the stabilizer group of a code, $\mathcal S = \mathcal Z(\mathcal G)$ ( \autoref{fig:algebras}), which describes the (abelian) symmetries of the circuit. We use both notions in this paper; where there is potential for ambiguity,  we will explicitly refer to code stabilizers and state stabilizers, but we may omit these qualifiers when the meaning is clear from context.]

We can  compute the dynamics of a state $\rho$  specified in this manner under projective measurements of operators from the Pauli group\footnote{More generally, one can efficiently classically simulate the Heisenberg evolution under any set of operations that preserves the Pauli group, but in this work we restrict ourselves to measurement-only dynamics.} as follows. 
Given a state $\rho$ specified by $\mathcal{S}(\rho) =\expval{s_1, s_2, \ldots, s_m}$ and a measured  Pauli  operator $g\in\pauligroup$,  our task is to give a prescription for the post-measurement state $\rho'$ [or equivalently $\mathcal{S}(\rho')$]. To that end, we distinguish three cases:
\begin{enumerate}
\item If $\pm g \in \mathcal{S}(\rho)$ so that $g$ is an element of the stabilizer group of $\rho$, then the state is unchanged by measurement, so that $\rho' =\rho$ and hence $\mathcal{S}(\rho) = \mathcal{S}(\rho')$, and the measurement outcome is the eigenvalue of $\rho$ with respect to $g$.
\item If $\pm g\notin \mathcal S(\rho)$, but if $g$ commutes with every generator $s_i\in S(\rho)$, 
then we simply add $g$ to $\mathcal{S}$, i.e. the post-measurement state is generated by the stabilizer set $S(\rho') = \expval{\pm g, s_1, s_2, \dots s_m}$.
\item Finally, if $\pm g\notin \mathcal{S}(\rho)$ and $g$ anticommutes with some $s_i\in \mathcal{S}(\rho)$, we can without loss of generality  assume that $g$ anticommutes with exactly one generator $s_1$, since we can simply redefine $s_j \to s_1 s_j$ for each $s_j\neq s_1$ that also anticommutes with $g$. In this case, we remove $s_1$ from the stabilizer group and replace it with $g$, so that $S(\rho') = \expval{\pm g, s_2, \dots s_m}$.
\end{enumerate}
In both cases (2) and (3), the sign of $g$ in $\mathcal S(\rho)$ is given by the measurement outcome.

It is worth briefly commenting on the nature of such dynamics for stabilizer and subsystem codes. In the stabilizer case, since the measured operators always commute, Rule 3 above is never invoked and the state stabilizer group can grow but is otherwise fully static. In contrast, in a subsystem code where the measured check operators need not commute, adding new state stabilizers can remove existing ones. A special case occurs when a state stabilizer coincides with one of the code stabilizers, whereupon --- as it then by definition commutes with all the check operators --- it is never removed from the state stabilizer group. Thus the measurement only dynamics of subsystem codes can be viewed in terms of a gradual accumulation of code stabilizers into the state stabilizer group as time evolves; this will be central to our discussion of the slow dynamics below.

\subsection{Entanglement-based characterization of dynamical states}

As noted by earlier studies~\cite{lavasani2021topological_2D, kitaevMO2023a,kitaevMO2023b,zhu2023kitaev_color_code,sharma2023BaconShor,klocke2022XZZXcode}, the ultimate fate of the measurement-only dynamics for $t\to \infty$ and any finite $N$ is for the system to reach a density matrix that is invariant under the  stabilizer group of the code.  In most models of interest, any stabilizer generator can be measured effectively by measuring a certain sequence of check operators. This is also the case for all codes considered here, although we also present a simple example where this is not possible in \appref{app:yao-kivelson}. Logical operators cannot be measured by measuring products of check operators; hence, the entanglement between the logical subsystem and the environment is an invariant of the dynamics.\footnote{The easiest way to see this is to note that  any state $\rho$ with nontrivial entanglement between the logical subsystem and the environment can be purified into a state on system and environment whose  stabilizer group  involves some number of operators that are products of logicals and operators on the environment. Since all measurements commute with the logicals (and trivially with environment operators) and logicals are themselves never measured, these operators always remain part of the state stabilizer group and hence the logical-environment entanglement is preserved.}

However, the approach to this late-time state can involve  several distinct dynamical regimes whose timescales can diverge in the thermodynamic limit, as we discuss in detail the next section. As we will demonstrate, these regimes can be  understood  by characterizing the  entanglement properties of the system. When considering how the measurements purify initially mixed states, the relevant notion of entanglement is that of the system with the environment, as captured by the  von Neumann entropy $S(t)= - \tr \rho (t) \log_2 \rho(t)$. On the other hand, when characterizing how measurements  generate entanglement in initially unentangled pure states, we instead focus on the bipartite entanglement entropy;  given a bipartition of the system into a subsystem $A$ and its complement $\bar{A}$, this is given by $S_A(t) =  -\tr \rho_A(t) \log_2 \rho_A(t)$, with $\rho_A(t) \equiv \tr_{\bar{A}}\rho(t)$. In both cases we  will be interested in the scaling of the respective entropies with (sub)system size.

In the case of stabilizer states [\autoref{eq:stab-rho}], the von Neumann entropy is given by the difference between the number of qubits and the number of independent generators $\mathcal S(\rho)$ of the stabilizer group, while the entanglement entropy $S_{A}(t)$ is given by the minimal number of generators that cannot be supported on only either $A$ or $\bar A$~\cite{fattal2004entanglement_stabilizer_formalism}. The latter computation involves some linear algebra in order to perform the requisite minimization procedure on the state stabilizer generators.

We study purification dynamics by considering near-maximally mixed initial states with $S(0) \propto N$ that are highly entangled with the environment;  at late times we expect conversely that $S(t)$ is lower-bounded by $K$, the number of logical qubits, with the exact value as $t\to \infty$ controlled by the details of the initial state.  Scalings intermediate between these arise can arise during the evolution owing to the interplay of the code structure with the locality of measurements, and require additional analysis. 

Turning to the scaling of the entanglement entropy of a subsystem, we employ the usual practice of distinguishing between volume law ($S_A\propto |A|$) and area law ($S_A\propto |\partial A|$) scaling of the dominant contribution, and between topological and trivial subleading contributions in area-law phases. Since when exploring scrambling we typically begin with a product state, $S_A =0$ initially and approaches either a volume or area law at late times.

Area law states can in turn be distinguished by subleading contributions to $S_A$. For instance, in a two-dimensional topologically ordered phase we have
\begin{equation}
    S_{A} = c \abs{\partial A} - \gamma,
    \label{eq:tee}
\end{equation}
where $\abs{\partial A}$ is the size of the perimeter of the region $A$, $c$ is some non-universal constant, and $\gamma$ is called the topological entanglement entropy (TEE)~\cite{kitaev2006topological,levin2006topological}. In our analysis of intermediate-time behaviour of measurement-only dynamics of subsystem codes, we will find  additional, non-local area-law contributions to the entanglement entropy that are leading (subleading) in spacial dimension $D=2$ ($D=3$) that arise as a consequence of their special symmetry properties.

Note that in probing ``generic'' scrambling behaviour it is important to choose a random initial state. This is because special initial states, such as product states in the local $Z$ basis, can be eigenstates of non-local stabilizer generators of the code and hence evade slow  dynamics linked to the formation of these stabilizers.

\section{Results\label{sec:results}}

\subsection{Overview}
We begin by giving an overview of our results, that we obtain from  different examples of subsystem-code circuits  with non-locally generated stabilizer groups. Each of these circuits illustrates a distinct aspect of the measurement-only dynamics, and as we discuss in detail in subsequent sections. A summary of the models that we study are shown in \autoref{fig:intro} (d).

\subsubsection{Slow Purification and Scrambling: ``Learning'' an Unknown State and the Role of Stabilizer Geometry\label{sec:slow-mode}}

\begin{figure}
\includegraphics{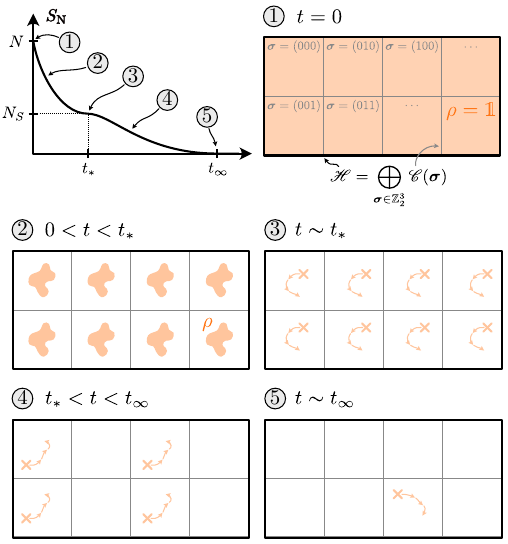}
\caption{
	Purification dynamics as learning of the state in Hilbert space. Assume for simplicity that all stabiliser generators of the code purify slowly (in the figure, $N_S=3$) and that there are no locical qubits ($K=0$). The Hilbert space $\vecspace H$ is the direct sum of subspaces $\vecspace S$, each labelled by a list of stabilizer eigenvalues $\vec \sigma \in \mathbb{Z}_2^{N_S}$.
	Purification dynamics then proceeds as follows: at time $t=0$, the state is maximally mixed $\rho = \id$, i.e. we have no information about the state [panel(1)]. 
	As check operators of the code are measured, we gradually learn about the state [panel (2)] and by the time $t_*$ we have typically learned everything about the state \emph{except} the stabilizer subspace it is in [panel (3)]. 
	Note that since the check operators do not commute, the state has dynamics within each stabilizer subspace on $\order{1}$ time scales [arrows in panel (3-5)].
	On a much longer time scale $t_\infty$, we also learn about the subspace that the state is in, by effectively measuring code stabilizer generators. In panel (4) for example, only the second generator has been measured with outcome $\sigma_2 = 0$.
	Finally after a typical time $t_{\infty}$, all stabilizer eigenvalues have been measured and the state is fully purified. 
	The state continues to have dynamics within the subspace, but cannot leave the subspace, i.e. the value of $\vec \sigma$ is a constant of motion.
}
\label{fig:purification_sketch}
\end{figure}

As we have noted, the study of measurement-only dynamics motivates the consideration of two quite different processes: purification and scrambling. In exploring purification dynamics, we initialize the system  in a maximally mixed state (so that the state stabilizer group is initially empty) and examine the time evolution of the von Neumann entropy $S(t)$. For an $[[N,K,D]]$ code and an initially maximally mixed state recall that we expect that $S(0)=N$, the number of qubits, and that $S(t \to \infty) = K$, the number of logicals, since the associated degrees of freedom can not be purified by measuring check operators alone. 
For scrambling dynamics in contrast the initial state is taken to be a trivial product density matrix, usually with $N$ single-qubit stabilizers. When not further specified, we randomly choose a single qubit Pauli $X_j$, $Y_j$, $Z_j$ independently for each site $j$ with probability $1/3$. 

At late times we expect that the density matrix will be invariant under the full  stabilizer group of the {\it code}, since such density matrices are fixed points of the dynamics. Starting from any initial state, code stabilizers will be added to the state stabilizer group when a specific sequence of check operators is measured. Once a code stabilizer is added to the state stabilizer group, it will not be subsequently removed since by definition it commutes with all measurements performed on the state.

Given that the late-time state is well-understood, the natural open question about the dynamics is to characterize the approach to this state, which is closely linked to the time scale on which the stabilizer generators are measured. 
As was observed in Refs. \onlinecite{kitaevMO2023a,kitaevMO2023b}, a local stabilizer generator will typically be measured in constant time $\tau_{*}\sim\order{1}$, whereas to measure a general stabilizer generator with support $Q$ can take takes a time exponentially long in the size of this support $\tau_{\infty}\sim e^{\order{\abs{Q}}}$. 
Such potentially long time scales make these natural candidates for generating ``slow'' dynamics and motivates our interest in models all of whose stabilizer generators are non-local in the sense of the preceding subsection, namely that they have support on some $\Delta$-dimensional subsystem. 

We emphasize that in order to have slow  dynamics at intermediate times it is important to have non-local stabilizer \emph{generators} rather than either non-local logical operators or local stabilizer generators, since the former cannot be measured by dynamics that preserve the code structure while a stabilizer that is a local product of check operators is quickly measured in $\order{1}$ steps. 
 In contrast to the measurement-only Kitaev model considered in Refs. \onlinecite{kitaevMO2023a,kitaevMO2023b} which has a finite number of non-local stabilizer generators, here we consider circuits where the number of non-local stabilizer generators $N_S$ grows with system size ($N_S \sim\order{L^{D-\Delta}}$). We  study cases with both string-like $\Delta=1$ and sheet-like $\Delta=2$ generators. 

As the relative rates of non-commuting check measurements are varied, whether a given non-local stabilizer generator is in fact a slow degree of freedom is dictated by its geometry. 
We find that for string-like stabilizer generators of a code, some subset of the generators is always measured quickly, the concrete subset being determined by the rate of competing check measurements. In contrast, in circuits with sheet-like stabilizer generators, for some relative rates of competing check measurements, \emph{all} generators may take an exponentially long time to be measured for. 
Here, by {\it competing} check operators we mean those that anti-commute with some intermediate state of the sequence of check measurements necessary to measure the stabilizer generator in question.  
The difference in ``stability'' between string- and sheet-like stabilizer generators may be intuitively understood by estimating the number of competing check operators, as we detail in \autoref{sec:membranebvstrings}. 
While the above result might seem technical, one can gain an intuitive understanding by considering the dynamics of  purification, within a fixed trajectory of measurement outcomes, as the process of learning information about an initially unknown state of the system by means of repeated measurements.
This is illustrated in \autoref{fig:purification_sketch}, where we show how the uncertainty about the state in the Hilbert space (indicated by orange shading) evolves in time during purification. For simplicity, in the figure we assume that all  $N_S = 3$ stabilizer generators are slow and that there are no logical degrees of freedom, so that $K=0$.
Starting from zero knowledge, $\rho=\id$ at $t=0$ in panel (1), measuring the outcomes of local measurements quickly fixes the state within each stabilizer subspace $\vecspace C(\vec\sigma)$ (indicated by gray lines in the figure), but does not, in general, determine {\it which} subspace the state is in. The typical time for this `local purification', i.e. for the state to learn everything except the code stabilizer eigenvalues is $t_*$, as shown in panel (3). Note that even after $t_*$, the state has dynamics as indicated by the arrows in panel (3), but remains pure within each subspace $\vecspace C(\vec\sigma)$. 
Local measurements learn code stabilizer generators   at a much lower rate ($\tau_\infty^{-1}$), which gradually fixes the state [panel (4)] until it is fully purified at a typical time $t_\infty$ shown in panel (5). 
Even at the longest times the system has dynamics within its final code stabilizer subspace, but the subspace remains fixed.

\subsubsection{Nonlocal Area Laws}

For circuits based on codes, the steady-state entanglement entropy can receive nonlocal contributions either from logical operators, observed e.g. in the toric code~\cite{lavasani2021topological_2D}, or from non-local stabilizers as was observed in the measurement-only Kitaev model~\cite{kitaevMO2023a,kitaevMO2023b}. 

While in both of these cases, the nonlocal contribution is topological (in the sense of \autoref{eq:tee}) and does not scale with system size, a key innovation in our work is to consider circuits where \emph{all} generators of the stabilizer group are necessarily non-local and take the form of subsystem symmetries. Specificially, they act nontrivially on a $\Delta < D$ dimensional subsystem, where $D$ is the full spatial dimension of the circuit.

The number of independent generators 
then grows subextensively with the codimension of the stabilizers, i.e. as 
$\order{L^{D-\Delta}}$. As we show in \autoref{sec:results}, if $\Delta=1$ so that the stabilizer generators each act on a line-like subsystem, they give a finite area-law contribution to the entanglement entropy of the steady state in any dimension $D\geq 2$. This contribution is non-local and hence at odds with the common wisdom that area law terms in the entanglement entropy stem exclusively from local contributions (similar to what is observed in Ref. \onlinecite{balasubramanian2023nonlocal} for the ground states of a class of local Hamiltonians). Note that $D=3$ codes admit the possibility of sheet-like stabilizers, but the corresponding non-local contribution is subleading to the area law as it scales as $\order{L}$.

Non-local stabilizers linked to subsystem symmetries  are also known to give rise to  spurious contributions to the ground state entanglement entropy in certain stabilizer codes~\cite{williamson2019spurious}. However, in the dynamical setting there is an important distinction between codes with stabilizer groups that cannot be locally  \emph{generated} and  those with stabilizer groups that contain nonlocal elements but nevertheless admit a local generating set. While both can lead to unusual entanglement scaling, for the reasons discussed above,
we do not in general expect slow measurement-only dynamics in cases where the stabilizer group can be generated by local operators.

\subsubsection{Frustration Graphs and Volume-Law Entanglement}
We also comment on a 
link between the late-time entanglement scaling and the characterization of a circuit in terms of its frustration graph, originally conjectured by Ref. \onlinecite{ippoliti2021MOD}.
Each node $j$ of the frustration graph corresponds to a check operator $g_j$; two nodes $i$, $j$ are connected by and edge if and only if the two corresponding checks do not commute, $[g_i, g_j]\neq 0$. The authors conjectured that a volume-law phase can  be generated by measurement-only dynamics only if the frustration graph of the circuit is non-bipartite. For the circuits based on subsystem codes considered in this work, the frustration graph is bipartite if the underlying code is of the Calderbank-Shor-Steane (CSS) form.  We  find that of the circuits we study only those based on non-CSS codes support volume law steady states, consistent with the conjecture of Ref.~\onlinecite{ippoliti2021MOD}. [Note, however, that while CSS code circuits always have bipartite frustration graphs, they are not the unique circuits with this property. Examples of non-CSS code circuits with bipartite frustration graphs  include the measurement-only toric code with single-qubit $Y$ measurements~\cite{lavasani2021topological_2D}, and the one-dimensional chain with $\{XX,YY,ZZ\}$ checks on nearest-neighbours, such that checks on even sites only anticommute with those on odd sites~\cite{ippoliti2021MOD}.]

\subsubsection{Measurement-only Analogue of Hilbert Space Fragmentation}

Given the equivalence of the algebraic structure of subsystem codes and the bond-algebraic formulation of Hilbert-space fragmentation, it is natural to ask  how fragmented Hilbert-space structure impacts dynamics that lack unitarity. While one can endow non-unitary time evolution (e.g. by imposing a similar block structure on Lindblad superoperators) with similar structure to the unitary case, extending the notion of fragmentation to the measurement-only setting has not to date been explored. 

First note that as long as we monitor the measurement outcomes, the measurement-only dynamics maps pure states to pure states. The resulting pure-state trajectories are then also restricted to only to exploring a fixed stabilizer subspace, and hence would resemble Panel (5) of \autoref{fig:purification_sketch}. 
This is analogous to the fragmented dynamics one would expect for unitary dynamics in a subsystem code, and hence might lead to similar signatures such as persistent oscillations in certain (spin-glass) observables. While systematic exploration of such aspects is likely to be interesting in its own regard, here we focus on the slow dynamics of \emph{entanglement}.
In this context our results show that, on imposing the additional caveat that we consider subsystem codes with non-local stabilizer generators, we can indeed define a type of `fragmented measurement-only dynamics' of scrambling or purification, where the system persists in a sub-maximally entangled or sub-maximally-purified state for a time scale that diverges (albeit sub-extensively) in system size. 
We conjecture that imposing this kind of additional structure beyond just the Hilbert-space decomposition of \autoref{eq:h-ssc} is in fact necessary for such slow dynamics of entanglement to arise in measurement-only models. Furthermore, to the best of our knowledge, no special role has been identified for nonlocally-generated symmetries in the context of fragmented unitary dynamics; it would be interesting to consider this in future work.

\subsubsection{Stability to Perturbations}

We also comment on a related but distinct problem: namely, the probability of (and hence the time scale on which) logical operators of error correcting codes are measured by single-qubit measurements. This been studied in a series in recent works~\cite{botzung2023robustness,lee2024randomly_monitored_codes}, which suggest a remarkable robustness of the logical subspace against local measurements.

A natural question is then to ask how robust the phenomenology discovered in this work (consisting of slow entanglement dynamics and the non-local late-time area laws) is to adding perturbations that break the symmetry of the circuit. We find that, as in most cases of Hilbert space fragmentation, neither the slow dynamics nor the non-local area law contribution persists upon introducing a finite probability of measuring single qubit operators --- at least in the examples considered here. As we now discuss, this  is a direct consequence of the nonlocal nature of \emph{all} stabilizer generators in our examples. 

Consider an environment initially entangled with the degree of freedom associated with the code stabilizer $S_j$. This means that the stabilizer group describing the state of the composite system contains generators of the form
\begin{equation}
    E_j = O^{(\rm env)}_j \otimes S_j
\end{equation}
but neither $O_{\rm env}$ nor $S_j$ alone.
When all measurements done on the system commute with $S_j$, then the only way to disentangle the environment is to measure $S_j$ itself, which takes time $\order{\exp L^\Delta}$. 
If we now perturb the dynamics by introducing a small density of measurements $M_k$ which do not commute with the $S_j$ however, then measuring the $M_k$ can also disentangle the system.

If there are both nonlocal and local code stabilizers, then the local stabilizers might be able to ``protect'' the nonlocal ones from local non-commuting measurements $M_k$. This is because in order to purify a nonlocal stabilizer generator, the operator $M_k$ has to commute with all local stabilizer generators of the code~\cite{botzung2023robustness}. Since the local stabilizer generators are measured with some finite rate (as is the case for example in the measurement-only Kitaev model~\cite{kitaevMO2023a,kitaevMO2023b}), there is then competition between adding operators to the state stabilizer group by  measuring stabilizers and removing them by measuring the $M_k$.

In contrast, the circuits considered in this work are based on codes where all stabilizer generators are nonlocal and hence are measured with a rate that is exponentially small in system size. In this case, since single-qubit measurements remove stabilizers with finite probability $p_s$, any $p_s > 0$ leads to purification in logarithmic time.  We have verified this numerically for the examples considered in this paper.

Note that the above argument is strongly reminiscent of the reasoning used to argue that the Bacon-Shor code does not have a error-correction threshold~\cite{eczoo_bacon_shor} and in particular no measurement threshold~\cite{lee2024randomly_monitored_codes}. It hence serves as another striking examples of the parallels between entanglement dynamics in circuits and  error correction (specifically, in this case,  the error correction properties of the underlying code).

\subsection{Warmup: The Ising Chain}
A particularly simple example of the above formalism is furnished by the projective 1D transverse-field Ising model\footnote{We thank Matteo Ippoliti for suggesting this illustrative example.}~\cite{LangBuchlerProjIsing}. This can be viewed as the measurement-only dynamics of a trivial subsystem code whose competing check operators are $g_i^{(X)}= X_i$, $g_i^{(Z)} = Z_iZ_{i+1}$, where $i=1, \ldots, N$ denote sites (and we identify the sites at $i=1$ and $i=N$), and $X/Z$ are Pauli operators. There is a single stabilizer $S_i^{(X)} = \prod_i g_i^{(X)}$ with $\Delta=1$, and no logical qubits so that $K=0$. If we  consider measurement-only dynamics in the `paramagnetic' which we measure the $g_i^{(X)}$ more frequently, we quickly measure the single stabilizer and the system purifies quickly. However if instead we consider the `spin glass' phase where we measure the $g_i^{(Z)}$ more frequently, then the system quickly learns all {\it local} information, but needs to perform a system-spanning sequence of $X$ checks in order to diagnose the global parity bit (encoded by the eigenvalue of $S_i^{(X)}$); since this is frustrated by measuring the $Z$ checks, there is an exponentially slow timescale for learning this single stabilizer and hence slow purification.

\subsection{The 2D Bacon-Shor Code\label{sec:2d_bsc}}

\begin{figure*}
\centering{}
\includegraphics{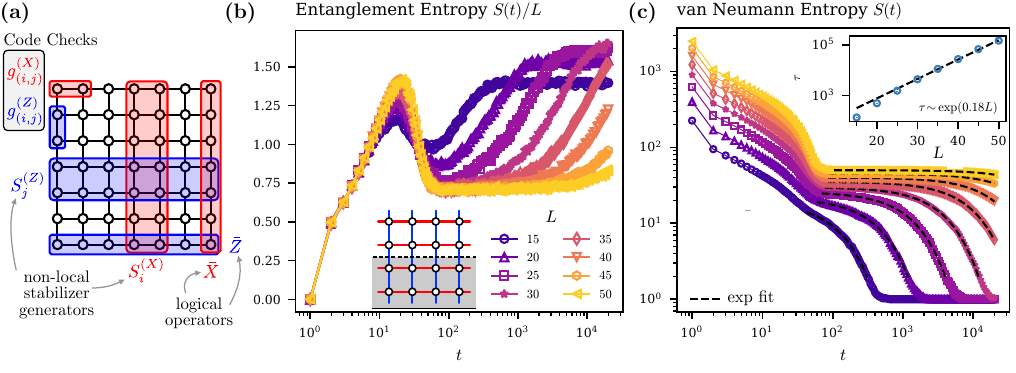}
\caption{
Entanglement Dynamics of the two-dimensional measurement-only Bacon-Shor circuit. (a) Summary of the circuit geometry. At each time step, we measure a random check operator $g_{ij}^{(X)}$, $g_{ij}^{(Z)}$ with probability $p_x$, $p_z=1-p_x$, respectively. The (Pauli-)symmetry group of the circuit is generated by the stabilizers $S_{i/j}^{X/Z}$ and logical operators $\bar X / \bar Z$, which can and cannot be written as a product of check operators, respectively. (b) Entanglement entropy across a cut in the $y$-direction (see inset), as a function of time for $p_x=0.49$ when starting from a trivial state. After an initial transient, the entanglement entropy settles to a value given by the local contribution to the area law, while the non-local contribution to the steady-state value $S_{\infty}$ builds only at exponentially long times. (c) Purification dynamics when starting from a fully mixed state. The inset shows a fit of the system-size dependence long time scale $t_\infty$.}
\label{fig:2d_bsc_highlights}
\end{figure*}

We begin our suite of examples by considering the best-known example of a code with only non-local stabilizer generators: the two-dimensional Bacon-Shor subsystem code~\cite{bacon2006bsc}. The geometry of the code is sketched in \autoref{fig:2d_bsc_highlights} (a). Qubits reside on the sites of a square lattice and a single two-body check operator is defined on each bond, depending on its orientation: 
\begin{align}
    g_{(i,j)}^{(X)} = X_{(i,j)} X_{(i+1,j)}, \quad
    g_{(i,j)}^{(Z)} = Z_{(i,j)} Z_{(i+1,j)},
    \label{eq:bsc2d_checks}
\end{align}
where $(i,j)$, $i,j=1\dots L$ denote the sites and $X/Z$ are Pauli operators. The stabilizer group of the Bacon-Shor code is generated by the so-called ``Bacon strips''\footnote{Pun intended, but not original to us.}, given by the product of $Z$ and $X$ checks along a given row and column, respectively
\begin{align}
    S_{i}^{(X)} = \prod_j g_{(i,j)}^{(X)}, \qquad
    S_{j}^{(Z)} = \prod_i g_{(i,j)}^{(Z)}.
    \label{eq:bsc2d_stabs}
\end{align}
Bare logical operators in turn are given by products of $Z$ and $X$ operators along a \textit{single} row and column, respectively:
\begin{align}
    \bar X = \prod_j X_{(0, j)}, \qquad
    \bar Z = \prod_i Z_{(i, 0)}.
\end{align}
Note that the choice of row/column of the logical operator is arbitrary, since different choices are related by multiplication with a code stabilizer.

In \autoref{fig:2d_bsc_highlights} (b), we show the half-system entanglement entropy $S(t)$ as a function of time for stochastic measurement-only dynamics (see \autoref{sec:stochmeasdyn} for details) starting from a product state. The data clearly reveals the existence of two dynamical regimes at long times: after an initial transient, the system settles to a local area law, while the non-local area law state that we expect for $t\to\infty$ is reached only after a time exponentially long in the linear system size $L$.

\begin{figure}
\includegraphics{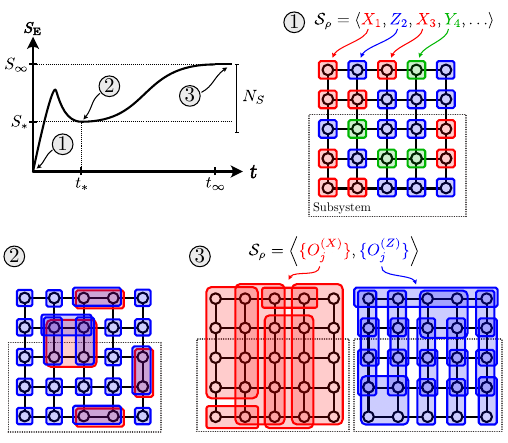}
\caption{
Scrambling in the 2D Bacon-Shor Code for $p_z > p_x$.
We sketch a typical set of state stabilizer generators for three characteristic times. 
(1) At $t=0$, the system is in a product state and the state stabilizer group is generated by randomly-choosen single-qubit Pauli operators. 
(2) At $t\sim t_*$, the state stabilizer group can still typically be generated by only local operators, but will no longer be an exact product state (and hence will generically have area-law entanglement).
(3) We now draw $X$ and $Z-$type state stabilizer generators separately for better visibility. The state stabilizer group at late times $t\sim t_{\infty}$ typically does not admit a local generating set of operators.
}
\label{fig:scrambling_sketch}
\end{figure}

We can rationalize this behaviour and our  characterization of  the entanglement at $t\sim t_*$ and $t\sim t_\infty$  as local and nonlocal respectively by considering the behaviour of the typical state stabilizer generators at these these characteristic times.  A representative sketch of these generators is shown in \autoref{fig:purification_sketch}, for the case $p_z > p_x$, which we now discuss.  The system is initialized in a product state at $t=0$, whose  state stabilizer group is generated by randomly chosen single-qubit operators.  As time proceeds and we measure check operators, the state stabilizers are no longer single qubits. Since the check operators in different directions do not commute, however, $Z$- and $X-$type state stabilizers grow along different axes, but their growth can be truncated by competing measurements; however, {once} a state stabilizer has grown to span the system it {coincides with} a code stabilizer and is thereafter invariant under the dynamics. 

To understand the consequences for entanglement, recall that the computation of the bipartite entanglement entropy $S_E$ follows by choosing state stabilizer generators such that a minimal number of them straddle the entanglement cut~\cite{fattal2004entanglement_stabilizer_formalism}. Evidently, if a code stabilizer is also part of this minimal generating set, this leads to a nonlocal contribution to $S_E$. However, whether this is the case  depends on the nature of the typical local stabilizers. (Since these are all state stabilizers, for conciseness we omit that qualifier below.) In the steady state of our measurement-only dynamics, this is controlled by the relative probability of $Z$- and $X-$type measurements. For $p_z>p_x$, $\mathcal{S}(\rho)$  typically contains more local $Z-$type stabilizers than $X$-type ones. In this case, although a nonlocal $Z$-type code stabilizer $S_j^{(Z)}$  will also be in $\mathcal{S}(\rho)$ with high probability, 
there are typically many local $Z$-type stabilizers that can be used to generate it; so any minimal generating set of $\mathcal{S}(\rho)$ is unlikely to require nonlocal generators in order to generate $S_j^{(Z)}$.
Conversely, a nonlocal $X$-type code stabilizer $S_i^{(X)}$  is {\it unlikely} to be in $\mathcal{S}(\rho)$ at early times since it requires a rare event (the probability is exponentially small in linear system size) but once created it remains in 
$\mathcal{S}(\rho)$ for all later times. Since at any given time there are typically very few local $X$-type stabilizers in $\mathcal{S}(\rho)$  that could generate such an $S_i^{(X)}$, it is highly likely that any generating set of $\mathcal{S}(\rho)$ will require nonlocal elements\footnote{Note that we cannot guarantee that any {\it specific} nonlocal code stabilizer is in the generating set, since it is nonunique.}. Thus the $S_i^{(X)}$s, once measured, will with high probability yield a nonlocal contribution to $S_E$.

This results in the following  picture of stabilizer dynamics for $p_z>p_x$. We begin in a product state, and as noted we start building local entanglement by measuring check operators. After an initial transient, at a time $t\sim t_*$ the state stabilizer group can still typically be generated by local (but generically not  \emph{single-qubit}) operators, and hence the subsystem entanglement entropy follows the usual area-law scaling with subsystem size. Furthermore 
the $S_j^{(Z)}$  have with high probability been effectively measured already. However, as noted above, they do not typically yield a nonlocal contribution to $S_E$. 
As the system evolves, an occasional rare event will create an  $S_i^{(X)}$; as noted above this will with high probability require another nonlocal generator for $\mathcal{S}(\rho)$ and hence yields a nonlocal contribution to $S_E$ (for a cut along $\vec{\hat x}$-direction). Each such rare event increases the nonlocal piece of $S_E$ by one bit, until a time $t\sim t_\infty$ --- corresponding to when all $N_S\sim L$ of the $S_i^{(X)}$ have been measured --- the entanglement reaches its saturation value $S_\infty$, {the nonlocal contribution to} which scales with the extent of the subsystem along the $\vec{\hat x}$ direction. 

The behavior for the case $p_x > p_z$ is equivalent, but with the roles of the stabilizers $S_j^{(Z)}$ and $S_i^{(X)}$ reversed, and the nonlocal contribution to $S_E$ given by the extent of the subsystem in the $\vec{\hat y}$-direction. Evidently, in both cases, there will be no slow buildup of entanglement for cuts that are not straddled by the slow stabilizers, since this direction does not have a nonlocal area-law contribution.

The dynamics can also be probed by studying the process of purification, i.e. by tracking the von Neumann entropy of the mixed state $\rho$ as a function of time when starting from a fully mixed initial state. We show this in \autoref{fig:2d_bsc_highlights} (c), where again two dynamical regimes are clearly visible: the system quickly [in a time of $\order{\log N}$] purifies down to an entropy $S = L-1$, and then decays slowly towards the infinite-time limit $S \to 1$ (which is set by the fact that the  logical qubit cannot be disentangled from the environment by this choice of dynamics). To verify this picture quantitatively, we fit the von Neumann entropy $S(t)$ at  times after  the initial  transient to an exponential decay, $S-1 \propto e^{-t/\tau}$, which defines a purification time scale $\tau$. We show the system-size dependence of $\tau$  in the inset of \autoref{fig:2d_bsc_highlights} (c), and find that it is well-described by an exponential, $\tau \propto e^{0.18 L}$.

The purification dynamics also shows two distinct dynamical phases as a function of the relative measurement rates $p_x$ and $p_z$. For $p_z>p_x$ (the case discussed above), the $Z$ stabilizers, despite their large support, are measured quickly whereas the $X$ stabilizers take an exponentially long time (in $L$) to be measured. For $p_z < p_x$ the roles of $X$ and $Z$ stabilizers are reversed. Finally, for $p_x = p_z$ the system shows critical dynamics: there is no longer a separation of purification timescales between $X$- and $Z$-type stabilizers, and instead the system shows algebraic scaling of the purification time, $S(t) \sim \exp(-t/L^z)$ with $z=3/2$~\cite{sharma2023BaconShor,lavasani2021topological_2D}.

\subsection{Going 3D: Membranes vs. strings in the 3D Bacon-Shor and CSS-Plaquette Code}\label{sec:membranebvstrings}

In the preceding section, we demonstrated the unusual scrambling and purification dynamics en route to a non-local late-time area law state in a two-dimensional model model with string-like non-local stabilizer generators.
A natural next step is to clarify the role played by the geometry of the stabilizer generators in this problem:- for instance, if the non-local generators were sheet-like rather than string-like. In order to investigate the effect of such changes to the geometry, we are led to consider three-dimensional subsystem codes. 

\begin{figure*}
\includegraphics{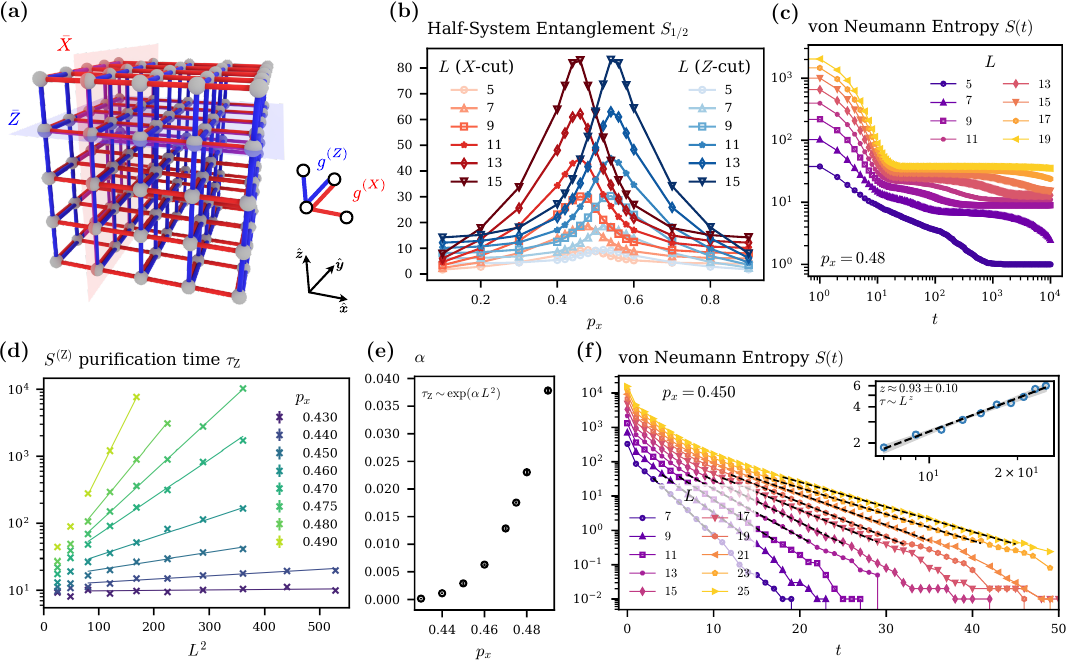}
\caption{
Entanglement dynamics of the 3D Bacon-Shor code. 
(a) Geometry of the check operators in the 3D Bacon-Shor code. 
(b) Half-system entanglement entropy along an cut in in the $x$-direction (red) and $z$-direction (blue). The two curves have pronounced maxima at different measurement rates $p_x$, indicating two distinct entanglement transitions.
(c) von Neumann entropy $S$ as a function of time $t$ for linear different system sizes $L$, in the intermediate phase $p_x=0.48$.
(d) purification time scale of the more frequently measured stabilizer (here $S^{(Z)}$) as $p_x$ is swept across the transition point at $p_x\approx0.45$. 
(e) growth rate of the fast stabilizer purification time scale, $\alpha$, as a function of measurement rate $p_x$.
(f) purification dynamics at the critical point, starting from an initial state where all code stabilizers have been measured already.
} 
\label{fig:3d_bsc_purification}
\end{figure*}

Concretely, we study two such codes: (1) the three-dimensional Bacon-Shor code~\cite{bacon2006bsc} with sheet-like stabilizer generators ($\Delta = 2$) and (2) the CSS-Plaquette code which has string-like stabilizer generators ($\Delta = 1$). We begin by discussing the former, which is a generalization of the two-dimensional case and sketched in \autoref{fig:3d_bsc_purification} (a). Qubits reside on the sites of a simple cubic lattice and checks are defined for bonds as follows: we define one $Z$($X$)-type check for each bond in the $\hat z$($\hat x$)-direction and \emph{both types} of checks for bonds in the $\hat y$ direction
\begin{equation}
    \mathcal G = \expval{
        X_{\vec r}X_{\vec r + \hat{\vec x}},
        X_{\vec r}X_{\vec r + \hat{\vec y}},
        Z_{\vec r}Z_{\vec r + \hat{\vec y}},
        Z_{\vec r}Z_{\vec r + \hat{\vec z}};\vec r
        }.
\end{equation}

Restricting ourselves to odd $L$\footnote{For even $L$, the three-dimensional Bacon-Shor code still has non-local stabilizer generators, but has no logical qubit (see e.g. the discussion in Sec. IX B of Ref. \onlinecite{brown2016quantum_memory_finite_temp}).}, the stabilizer generators of this code are given by generalizing ``Bacon strips'' to ``nailbeds'': products of checks along planes perpendicular to the $\hat z$ and $\hat x$ direction, respectively. 
The logical operators are products of single-qubit Pauli operators along the same set of planes [see $\overline X$ and $\overline Z$ indicated in \autoref{fig:3d_bsc_purification} (a)]. Similar to the two-dimensional case, the choice of plane here is arbitrary, since products of single-qubit paulis along any give plane are related by multiplication with stabilizers. The three-dimensional Bacon-Shor code hence encodes a single logical qubit.

The presence of nonlocal, sheet-like stabilizer generators again leads to slow scrambling and purification dynamics, yet with the important difference that now both sets of sheet-like stabilizers simultaneously show slow purification for some range of the measurement rate $p_x$. 
To demonstrate this, we show first in  \autoref{fig:3d_bsc_purification} (b) the half-system entanglement entropy along both a cut perpendicular to the ${\hat x}$ (red) and perpendicular to the ${\hat z}$ direction (blue), as a function of the measurement rate $p_x$ of $X$-type check operators.
Note that each choice of direction will cut a particular type of code stabilizer and not the other.
The two curves show a maximum diverging with system size at two distinct measurement rates, suggesting two distinct entanglement transitions at $1- p_{c, z} = p_{c, x} \approx 0.45$, related two the two distinct types of code stabilizers.
To study the purification dynamics in the three phases, we show first, in panel (c), the total von Neumann entropy of the system as a function of time $t$, in the intermediate phase at $p_x = 0.48$. The entropy clearly exhibits two plateaus whose width grows with system size, and its late-time behavior is indeed fit by a functional dependence of the form
\begin{equation}
    S(t) = 1 + A \, e^{-t/\tau_Z} + B\, e^{-t/\tau_X},
\end{equation}
where both $\tau_Z$ and $\tau_X$ grow exponentially with the square of the linear system size.
This is shown explicitly in \autoref{fig:3d_bsc_purification} (d), where we plot the faster of the two time scales ($\tau_Z$ since $p_z > p_x$) as function of system size $L^2$ for a range of measurement rates $p_x$ both below and above the critical rate $p_{c, x}$. The solid lines are fits to
\begin{equation}\label{eq:3DBStau}
    \tau_Z \sim \exp(\alpha L^2)
\end{equation}
where the growth rate $\alpha$ itself  depends strongly on the measurement rate $p_x$, as shown in \autoref{fig:3d_bsc_purification} (e). In particular, the growth rate extracted from our numerics vanishes slightly below the critical rate $p_{c, x}$, indicating fast purification of the $Z$-type code stabilizer. We attribute the disagreement between of the point where the growth rate vanishes, and the critical rate extracted from the half-system entanglement entropy, to finite-size effects.

Finally, in panel (f) we show the critical purification dynamics at $p_x = p_{x, c}$, starting from a state where all code stabilizers have been measured (but nothing else), which isolates the critical scaling from the slow purification of the $Z$-type stabilizer. As expected, the purifaction time scales as a power law in system size, with the dynamical exponent fitted as $z = 0.93\pm 0.10$.

In summary, our numerical results suggest that the three-dimensional Bacon-Shor code has two distinct entanglement transitions, which can be diagnosed by considering the entanglement entropy across different cuts, and corresponding to the onset of slow purification dynamics for the two types of code stabilizers. Hence, in contrast to the two-dimensional case, where  either the $X$- or $Z$-type code stabilizers always purify fast, in three dimensions there exists an intermediate phase where both code stabilizers purify only on a time scale exponentially large in system size.

We attribute the differing measurement-only dynamics of the two- and three- dimensional Bacon-Shor codes to the contrasting geometry of the stabilizer generators. To see this, consider how stabilizer operators are measured in  stochastic measurement-only dynamics. Viewing the dynamics in terms of the state stabilizer group, to measure a stabilizer with support on a $\Delta$-dimensional subsystem, one has to measure a sequence of $L^{\Delta}$ check operators. 
At any intermediate step, the state stabilizer group will contain  `incomplete' code stabilizers: these are either  open strings that would eventually grow into a string-like ($\Delta=1$) stabilizers, or  open membranes that would eventually grow into sheet-like ($\Delta=2$) stabilizers. Such a code stabilizer is measured (`completed') only when this open string or open membrane  spans the full system. Apart from the trivial outcome of no change, there are two possible fates for each incomplete code stabilizer in the subsequent time step: it can either grow, by measuring a commuting check operator, or shrink by measuring a non-commuting check operator, both located at its perimeter.
Our numerical results, in this language, suggest that similar to what is known for the fluctuations of magnetic domains in the one- and two-dimensional classical Ising model, membrane-like incomlete stabilziers are harder to grow against competing fluctuations in the bulk than their string-like counterparts. 
In fact, even biasing the measurement \emph{in favor} of their growth ($p_{x, z} > 0.5$) may not suffice to guarantee fast purification, opening up the possibility for an intermediate phase. 
We therefore conjecture that generically, it is the geometry of the stabilizers which dictates whether the entanglement and purification transition is unique, or whether there are two transitions, related by a Kramers-Wannier duality \cite{ippoliti2021MOD}, with an intermediate phase where \emph{all} stabilizer generators purify slowly. 

To test the above hypothesis, we also consider a three-dimensional model with string-like stabilizer generators, and confirm that in this case there are again only two distinct dynamical phases separated by a single critical point at $p_x=p_z$ (akin to the two-dimensional Bacon-Shor code). Naturally, this model also features a leading-order non-local contribution to the entanglement entropy in the area law phase, owing to the  $\order{L^2}$ string-like code stabilizers.

\begin{figure}[t]
\includegraphics{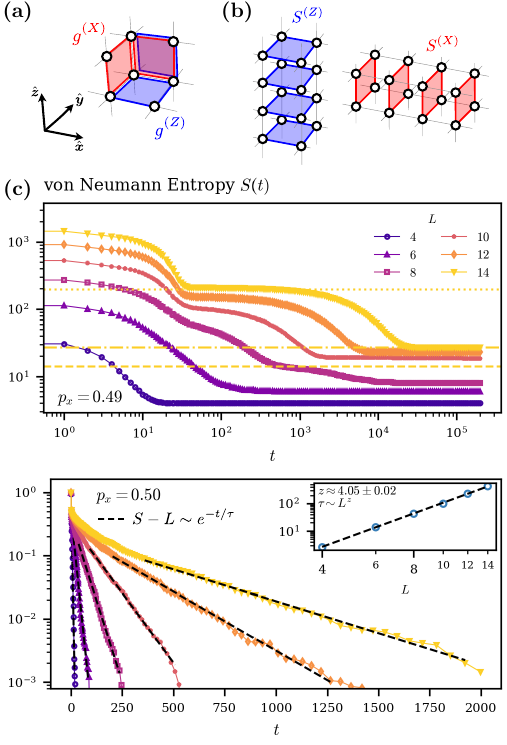}
\caption{Purification dynamics of the CSS plaquette code. (a) Geometry of four-body checks of the code. (b) geometry of string-like stabilizer generators. (c) von Neumann entropy as a function of time for the phase where Pauli-$Z$ stabilizers purify fast $p_x < p_z$ (top) and at the critical point $p_x = 0.5$ (bottom). The horizontal lines in the top panel are a guide to the eye (see main text for details). For the critical point, we plot the entropy density and fit the purification time $\tau$ at late times. The scaling of $\tau$ with system size is compatible with a power law with exponent $z=4$ (see inset). }
\label{fig:css_plaquette}
\end{figure}

This three-dimensional model is based on a code that we term the ``CSS plaquette code''. This is again defined on qubits situated on the sites of a simple cubic lattice, but the check operators now involve four qubits defined on the elementary plaquettes of the cubic lattice. As is shown in \autoref{fig:css_plaquette} (a), four-body Pauli-$X$ checks are defined on every plaquette perpendicular to the $\hat x$ and $\hat y$ directions, and four-body Pauli-$Z$ checks are defined on every plaquette perpendicular to the $\hat z$ and $\hat y$ directions.
The geometry of the full stabilizer group is quite involved and is different for odd and even system sizes (see \appref{app:plaquette_code} for details). Most importantly for the present purposes, the code has $\order{L^2}$ string-like stabilizer generators, sketched in \autoref{fig:css_plaquette} (b), which, per our conjecture above, should imply the existence of a single dynamics transitions, as well as a non-local contribution to the area law. This is demonstrated in \autoref{fig:css_plaquette} (c), where we show purification dynamics in the phase where the $Z$ stabilizers purify fast (top) as well as for the critical point (bottom). 
For $p_x \neq p_z$, the von Neumann entropy shows a clear plateau at $S \sim L^2$ (indicated by a dotted line for the largest system size), whose width grows exponentially with the linear system size $L$. There is also a second plateau at $S\sim L$ (indicated by a dash-dotted line, again only for the largest system size), whose width grows exponentially with the square of the linear system size $L^2$; this is only present for even system sizes for which the system also has sheet-like stabilizer generators. For odd system sizes, these sheet-like operators are logicals and hence they simply contribute to the infinite-time limit $S(t\to\infty)$. Finally, for $t\to\infty$ the entropy approaches the limit $S\to k$ (indicated by a dashed line), where $k=L$ is the number of logical qubits of the CSS-plaquette code for even $L$. 
For $p_x=p_z$, the system shows critical scaling of the purification time: at late times the data is compatible with
\begin{equation}
    S(t) \sim \exp(-\gamma \,t/L^z),
\end{equation}
with, remarkably, a subdiffisuive scaling $z\approx4$ [see bottom panel of \autoref{fig:css_plaquette} (c)].
This dynamical exponent is different both from that seen at the critical point of the two-dimensional~\cite{sharma2023BaconShor,lavasani2021topological_2D} ($z=3/2$) and three-dimensional Bacon-Shor circuit ($z=0.93\pm 0.10$ see above), as well as from those seen in the critical phase of the measurement-only Kitaev honeycomb model ($z=1/2$ at early times \cite{kitaevMO2023b} and $z=2$ at late times \cite{kitaevMO2023a, kitaevMO2023b}). 
In all other cases the $z \leq 2$ scaling is superdiffusive in contrast to the subdiffusive scaling seen here. As was suggested in Ref. \onlinecite{lavasani2021topological_2D}, such unusual exponents might be connected to the presence of subsystem symmetries. Drawing a precise connection between symmetry properties and dynamical exponents  remains an important topic for future work.
Finally, we note from the bottom panel of \autoref{fig:css_plaquette} (c) that there is no additional plateau associated with the sheet-like stabilizers at the critical point, suggesting that these purify in polynomial time. This demonstrate that while the geometry of code stabilizer generators plays an important role in determining purification dynamics and in particular the structure of the phase diagram, the exact relation is highly nontrivial.

\subsection{Adding Frustration: 3D Compass Code and non-CSS Plaquette Code}

We now turn to  measurement-only circuits with non-bipartite frustration graphs. Consistent with the connection originally conjectured in Ref. \onlinecite{ippoliti2021MOD}, we find that such circuits support a volume-law phase as part of their phase diagram. Specifically, we consider circuits based on  two different three dimensional non-CSS codes, each of which is obtained by modifying one of the CSS codes studied in the preceding section.

The first of these is a modification of the three-dimensional Bacon-Shor code, where as before check operators are defined as before on bonds of simple cubic lattice. However, we now impose Pauli-$X$, $Y$ and $Z$ checks on bonds in the 
$\hat x$, $\hat y$ and $\hat z$ directions, respectively:
\begin{align}
    g_{\vec r}^{(\alpha)} = \alpha_{\vec r} \alpha_{\vec r + \hat{\vec \alpha}},
    \label{eq:3d_compass_checks},
\end{align}
where $\alpha\in\{X, Y, Z\}$.
It is easy to verify that this subsystem code still encodes a single qubit, and has $3L-3$ stabilizers, which are products of $X$, $Y$ and $Z$ checks across planes perpendicular to their respective bond direction. In the following, we use the term ``three-dimensional compass circuits''  to refer to measurement-only circuits based on this code, since when considering the Hamiltonian $H=-J\sum_{g \in \mathcal G} g$, we obtain the so-called  ``three-dimensional 90$^\circ$ compass model''~\cite{nussinov2015compass_review}.

\begin{figure}[t]
\includegraphics{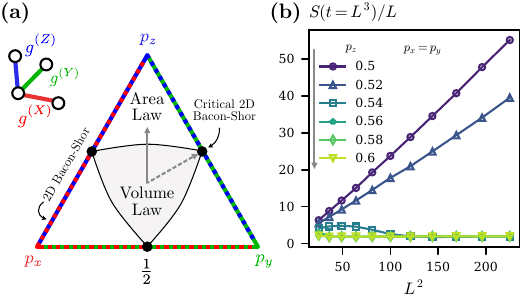}
\caption{
Conjectured entanglement phase diagram of the 3D compass circuit (see main text for details). (a) Check geometry and topology of the phase diagram. We numerically scanned the phase diagram along the lines indicated in gray. (b) von Neumann entropy after polynomial time as a function of measurement rate $p_z$ along the line $p_x = p_y$ indicated [solid gray line in panel (a)]. The data shows a clear transition from volume to area-law behavior at a critical rate of about $p_c \approx 0.54 \pm 0.02$.}
\label{fig:3d_compass_phases}
\end{figure}

The phase diagram that we conjecture based on studying purification dynamics along two cuts in parameter space is shown in \autoref{fig:3d_compass_phases}. A volume law phase occupies the center of the phase diagram, around $p_x=p_y=p_z$, and biasing the measurements towards one of these choices induces a transition into an area-law phase, as shown in panel \autoref{fig:3d_compass_phases} (b). 
This shows the von Neumann entropy after polynomial time $t=L^3$, when starting from a maximally mixed initial state. We expect that this number grows with the volume of the system $\order{L^3}$ in the volume-law phase~\cite{gullans2020mipt_purification} while it grows with the number of slowly-purifying stabilizers [here $\order{L}$] in the area-law phase. Indeed, plotting $S(t=L^3) / L$ against $L^2$, we see that the data shows linear growth for $p_z \lesssim0.52$,  it saturates to an number independent of system size for $0.52 \lesssim p_z$. 
We note that for the system sizes studied here, the fastest stabilizers $S^{(Z)}$ have already been measured at time $t=L^3$, so we observe  $S(t=L^3) = 2L - 1$ for $0.52 \lesssim p_z$.

We also numerically studied purification dynamics along the line $p_z=p_y$, shown as gray-dashed arrow in \autoref{fig:3d_compass_phases} (a), finding that the volume law phase does extend at least to $p_x =0.01$. We therefore conjecture that the system is in fact in a volume law phase for any $p_x > 0$. 
Along the boundaries of the phase diagram (i.e., where one of the three $p_i=0$), the system decouples (up to application of a local unitary) into $L$ independent copies of two-dimensional $L\times L$ Bacon-Shor code circuits.

Finally, we can similarly define a non-CSS version of the plaquette code that we introduced in the previous section (also see  \appref{app:plaquette_code} for details). The check operators of this ``non-CSS plaquette code'' are four-body Pauli-$X$, $Y$ and $Z$-type checks, all defined on the plaquettes on a simple-cubic lattice: instead of measuring both $X$ and $Z$ operators on plaquettes perpendicular to the $\hat y$ direction as in the CSS version, here we measure a four-body Pauli-$Y$ check with probability $p_y$. 
We have numerically computed purification dynamics along the same paths in parameter space as for the three-dimensional compass circuits [gray solid and dashed lines in \autoref{fig:3d_compass_phases} (a)]. Again, our numerical results suggest that the system undergoes a transition from a volume to area law phase along the line $p_z>p_x=p_y$ ($p_c\approx 0.51 \pm 0.02)$ and that it is in a volume law phase for $p_z=p_y > p_x > 0$. In contrast to the three-dimensional compass circuit case, the area law phases now have a non-local contribution to the entanglement entropy at leading order in addition to their distinct entanglement dynamics. Potentially, both models considered in this section could realize an even richer phase diagram than the one conjectured in \autoref{fig:3d_compass_phases} (a), but a detailed investigation of this is beyond the scope of the present work.

\section{Concluding Remarks\label{sec:conclusion}}

We  have presented the first systematic study of measurement-only dynamics in subsystem quantum error-correcting codes; for reasons described in the introduction such codes offer a particularly natural route to building code properties into circuit dynamics. In this context, our work singles out a special subset of codes whose stabilizer groups cannot be generated by purely local operators. Such codes are most conveniently constructed by considering models with nonlocally-generated subsystem symmetries: the symmetry generators are the nonlocal stabilizers, and their number diverges (subextensively) in the thermodynamic limit. 

The presence of such nonlocal stabilizers has striking consequences: 
\begin{itemize}
\item the  entanglement entropy along steady-state trajectories of the measurement-only dynamics of such codes acquires an unusual contribution linked to the nonlocal stabilizer generators. Strikingly, where the entanglement entropy of these states satisfies an area law, such nonlocal terms can contribute at leading order, in contrast to the conventional wisdom that area law contributions stem from short-range entanglement.
\end{itemize}

Furthermore, the approach to the late-time state is itself characterized by two distinct dynamical regimes, manifest in both scrambling and purification:
\begin{itemize}
\item in the scrambling dynamics of an initial product state, the local contribution to the entanglement saturates quickly, in $\order{\log L}$ time steps, whereas the non-local piece is only saturated in a time that is exponentially long in system size.

\item in the purification dynamics of an initially maximally mixed state, the system initially purifies relatively quickly up to a time of $\order{\log L}$ when all the local degrees of freedom have been disentangled from the environment or equivalently, all but the nonlocally encoded information about the state has been extracted from the system; thereafter, the rate of purification slows down and as the nonlocal stabilizers are purified, which again takes a time exponentially long in system size, leaving only the logicals (which cannot be measured) still entangled with the environment, or equivalently, their value remains unknown even to an observer that has access to the full history of measurement outcomes.
\end{itemize}

To establish and rationalize these results, we numerically simulated measurement-only dynamics in several different codes in spatial dimensions $D=2$ and $3$. In each of these models \emph{all} stabilizer generators of the code are nonlocal and involve symmetries that act on a subsystem of dimension $0< \Delta < D$, leading to a  non-local contribution to the entanglement entropy that scales  with the number of stabilizer generators, $N_S \sim L^{D-\Delta}$. 
We also showed that whether a given stabilizer generator is ``slow'', [meaning that the time scale on which it is measured diverges as $\tau_{\infty}\sim \exp(\gamma L^\Delta)$], in general depends on its geometry as well as the relative measurement rate of different check operators. 
To wit, we found that if all stabilizer generators are sheet-like with $\Delta = 2$ there may exists a dynamical phase in which all generators are measured slowly, with a rate that vanishes exponentially in the limit $L\to\infty$. In contrast, such a phase seems to be absent in models with string-like stabilizer generators.
We also found that in accord with a conjecture made in Ref. \onlinecite{ippoliti2021MOD}, circuits based on CSS codes do not support volume-law phases in their phase diagram, while those based on non-CSS codes do. 

Our results present a further step in the exploration of the dynamics of quantum many-body systems out of equilibrium. Perhaps most importantly, they are a clear demonstration that the algebraic structure of operators 
underlying a given dynamics (here, the operators measured in the measurement-only dynamics) can have striking consequences for both the steady-state entanglement properties as well as the  approaching to this state, even beyond the setting of unitary dynamics. We also drew a concrete parallel between the latter observation and the phenomenon of Hilbert space fragmentation in unitary dynamics, and its explication based on commutant algebras~\cite{moudgaya2022fragmentation}.

Multiple aspects of our work suggest directions worthy of future study. It would be fascinating to see to what extent our results can be generalized beyond stabilizer dynamics and the measurement-only setting. 
Staying within the context of measurement-only dynamics based on subsystem codes, first an important question for future work remains to understand the universality classes of the dynamical transitions, given that the critical behaviour is neither captured by existing mappings to percolation \cite{lavasani2021topological_2D} nor by those to loop models \cite{nahum_skinner2020majoranas,nahum2013crossings}. Second, there are other examples whose study could yield entirely novel phenomenology, such as circuits based on two-dimensional compass codes~\cite{li2019compass_codes}, where the size of stabilizer generators follows a broad distribution.
The direct parallel of our results to Hilbert space fragmentation suggests that all the codes studied here  may  be relevant in that context as well. Also, while the thermalization of subsystem codes under unitary dynamics has been studied previously~\cite{wildeboer2022inifite_temperature}, there is no systematic exploration of the difference between the unitary dynamics of systems with (subsystem) symmetry groups that can be generated by local operators (such as the parent Hamiltonian of the cluster state~\cite{williamson2019spurious}) and those that cannot (e.g. the 2D Bacon-Shor code).
Finally, the three-dimensional plaquette codes studied in this work have to the best of our knowledge not been previously studied, yet have favorable code parameters (see \appref{app:plaquette_code}). It may be of independent interest to explore their error-correcting properties even away from the dynamical setting considered here.

\begin{acknowledgments}

We are particularly grateful to Gurkirat Singh, for collaboration on related work and making us realize an error in an earlier version of this manuscript.
We  thank Matteo Ippoliti, Steven Simon, Victor Albert, Sarang Gopalakrishnan, and Vedika Khemani for valuable discussions and correspondence on related topics, and are especially grateful to Matteo Ippoliti for detailed comments on an earlier manuscript. 
B.P. acknowledges funding through a Leverhulme-Peierls Fellowship at the University of Oxford and the Alexander von Humboldt foundation through a Feodor-Lynen fellowship. 
S.A.P. acknowledges support from EPSRC Grants EP/S020527/1 and 	EP/X030881/1. 
\end{acknowledgments}

\appendix

\section{Measurement-only dynamics of the Yao-Kivelson Model\label{app:yao-kivelson}}

\begin{figure}
\includegraphics{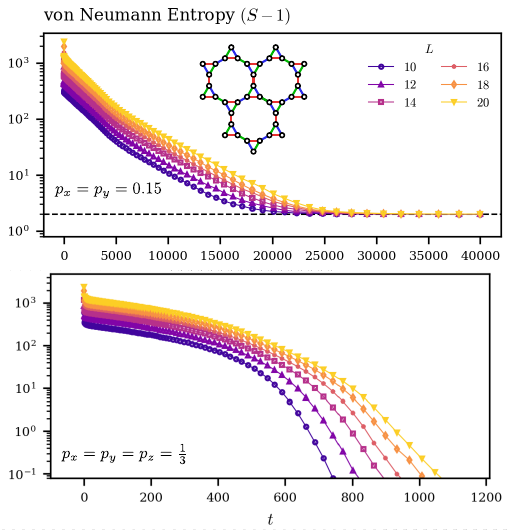}
\caption{Purification dynamics of the measurement-only Yao-Kivelson circuit. The geometry of the measured two-body operators is sketched in the top panel. The bottom panel shows von Neuman Entropy density $(S-1)/L^2$ as a function of time at the isotropic point $p_x=p_y=p_z$, which is compatible with critical scaling. The top panel shows the same quantity for strongly anisotropic measurement rates, compatible with a topological ordered phase.}
\label{fig:yao_kivelson}
\end{figure}

In this section we give an example of a case where not all \emph{stabilizer} generators of the code can be measured by a sequence of check measurements. The example is based on a Kitaev Hamiltonian on a decorated honeycomb lattice, first considered by Yao and Kivelson in Ref. \onlinecite{yao_kivelson2007}. The geometry of the lattice is shown in the bottom left of the top panel in \autoref{fig:yao_kivelson}: it is obtained from the honeycomb lattice by replacing each site with a triangle, which preserves its three-colorability. Here, we consider stochastic measurement-only dynamics analogous to that studied for the Kitaev model on the honeycomb lattice in Ref. \onlinecite{kitaevMO2023a,kitaevMO2023b}. That means given the three-coloring indicated in \autoref{fig:yao_kivelson}, at each time step we measure a two-body Pauli-$X$/$Y$/$Z$ operator on a randomly chosen red/green/blue bond with probability $p_x$/$p_y$/$p_z$ respectively.

Before discussing the numerical results in more detail, we flag perhaps most striking difference between the Yao-Kivelson circuit and the standard Kitaev circuit: namely, that the former does not purify the system even at infinite times. We find $S(t\to\infty) = 1$, whereas $S(t\to\infty) = 0$ for the Kitaev circuit on the honeycomb lattice. It turns out that this is a direct consequence of the presence of odd-length loops  (triangles) in the model, as we now discuss. 
First note that when viewed as a subsystem code, the Yao-Kivelson model has no logical qubit and the stabilizer group is given by products of checks along closed loops on the lattice. Hence, when we consider the system on a torus, the stabilizer group is  generated by products of checks around the elementary plaquettes of the lattice (triangles and octagons), as well as two non-local loops that wrap around the two inequivalent handles of the torus.
While even-length loops of check operators can be measured effectively by a sequence of check measurements (in the sense that after this sequence of measurements, the respective product will be part of the state-stabilizer group), this is not true for odd-length loops. In fact, it is straightforward to verify that, given three sites $i, j, k$, no sequence of measurements of the operators $X_i X_j$, $Y_j Y_k$ and $Z_i Z_k$ will effectively measure their product, i.e. the operator $Y_i Z_j X_k$.
For the Yao-Kivelson circuits this means that exactly one  stabilizer degree of freedom, associated with a single triangle, remains unpurified even for $t\to\infty$ when measuring only nearest-neighbor check operators. Note that one can effectively measure products around two triangles, so that only a single degree of freedom will remain mixed globally.

For completeness,  in \autoref{fig:yao_kivelson} we show the purification dynamics for both the isotropic point $p_x=p_y=p_z$ (bottom panel) and for strongly anisotropic measurement rates $p_x=p_y=0.15$, $p_z=0.7$ (top panel). 
The data is consistent with the two points being in distinct phases. At the isotropic point, the system in the observed time purifies all acessible degrees of freedom, i.e. $S\to 1$. For strong anisotropy, the nonlocal stabilizer generators purify slowly, leading to $S\to3$ in the observed time. These two behaviours are consistent with the system being in a critical and topologically ordered phase for $p_x=p_y=p_z$ and $p_x=p_y=0.15$, $p_z=0.7$, respectively. 

A detailed study of the two phases and the transition is beyond the scope of this appendix, but remains an important task for future work. In particular it would be interesting to study the consequences of the fact that the lattice here is not bipartite, which means that the Yao-Kivelson model presumably maps onto a classical loop model \emph{with crossings}~\cite{nahum2013crossings}. This in particular is different to the measurement-only honeycomb model, which can be mapped onto a classical loop model without crossings~\cite{nahum_skinner2020majoranas, klocke2023majorana},

\section{Detailed description of the Plaquette codes\label{app:plaquette_code}}

In this appendix, we provide some details about the CSS and non-CSS plaquette codes studied in the main text. 

Qubits for both codes reside on the sites of a simple cubic lattice and the check operators are four-body Pauli operators situated on plaquettes of the lattice. In the following, we give a detailed description of all check operators, stabilizers, and logical operators.

For the CSS codes, there are four check operators defined per lattice site $\vec r$
\begin{subequations}
\begin{align}
    g_{\vec r, \hat{\vec x}}^{(X)} &= X_{\vec r}\, 
                       X_{\vec r + \hat{\vec y}}\, 
                       X_{\vec r + \hat{\vec z}}\,
                       X_{\vec r + \hat{\vec y} + \hat{\vec z}},\\
    g_{\vec r, \hat{\vec y}}^{(X)} &= X_{\vec r}\, 
                       X_{\vec r + \hat{\vec x}}\, 
                       X_{\vec r + \hat{\vec z}}\, 
                       X_{\vec r + \hat{\vec x} + \hat{\vec z}},\\
    g_{\vec r, \hat{\vec z}}^{(Z)} &= Z_{\vec r}\, 
                       Z_{\vec r + \hat{\vec x}}\, 
                       Z_{\vec r + \hat{\vec y}}\, 
                       Z_{\vec r + \hat{\vec x} + \hat{\vec y}},\\
    g_{\vec r, \hat{\vec y}}^{(Z)} &= Z_{\vec r}\, 
                       Z_{\vec r + \hat{\vec x}}\, 
                       Z_{\vec r + \hat{\vec z}}\, 
                       Z_{\vec r + \hat{\vec x} + \hat{\vec z}}.
\end{align}
\label{eq:css_plaquette_checs}
\end{subequations}

\begin{figure}
\includegraphics{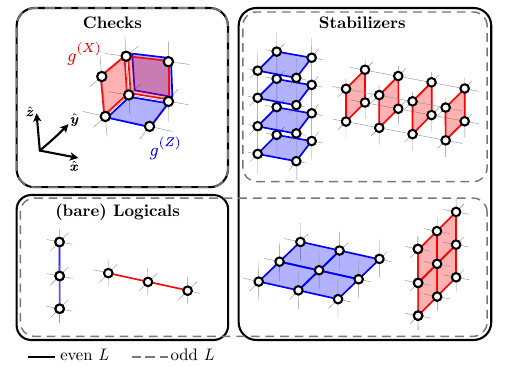}
\caption{Geometry of the CSS plaquette code. Checks are defined on the plaquettes of a simple cubic lattice of linear size $L$. For even $L$ (solid lines), there two kinds of stabilizers: one takes the form of string-like products of checks (top right) and another one takes the form of sheets of Pauli operators (bottom right). The bare logical operators (bottom left) take the form of strings of single-qubit Pauli operators. For odd $L$, the sheet-like stabilizer operators become logical operators.}
\label{fig:plaquette_codes_geometry}
\end{figure}

We sketch the geometry of checks, stabilizers and logical operators for both even and odd linear system sizes in \autoref{fig:plaquette_codes_geometry}.
On the top left of the figure we sketch the check operators also written in \autoref{eq:css_plaquette_checs}. Next to it, we show string-like products of checks, which can be written as 
\begin{equation}
    S^{(\alpha,\mathrm{str})}_{\vec r} = 
    \prod_{j=1}^L g^{(\alpha)}_{\vec r + j \hat{\vec \alpha}, \hat{\vec \alpha}}
\end{equation}
and which commute with all checks and are hence in the stabilizer group of the code. 
On the bottom left of the figure, we show string-like products of $X$ and $Z$ operators, which also commute with all check operators and hence are logical operators of the code
\begin{align}
    \bar X^{(\mathrm{str})}_{\vec r} = \prod_{j=1}^L X_{\vec r + j \hat{\vec x}} \qquad
    \bar Z^{(\mathrm{str})}_{\vec r} = \prod_{j=1}^L Z_{\vec r + j \hat{\vec z}}.
\end{align}
Finally, on the bottom right of the figure we show sheet-like products of Pauli operators, which always commute with all checks, but can be written as a product of checks only if the linear system size $L$ is even. This means, these operators are bare logicals for odd $L$ but stabilizer generators for even $L$.

Using the above, it is straightforward to see that the CSS plaquette codes on a $L\times L\times L$ simple cubic lattice with periodic boundary conditions are a family of codes with 
\begin{equation}
    [[N, K, D]] = \begin{cases}
        [[L^3, 3L-2, L]] & \text{if $L$ even}\\
        [[L^3, L, L]] & \text{if $L$ odd}
        \end{cases}.
\end{equation}
There are $2(L-1)^2$ independent string-like stabilizer generators if the system size is odd and an additional $4(L-1)$ independent sheet-like stabilizers if the system size is even.

For the non-CSS plaquette code we define three check operators per site $\vec r$ as
\begin{subequations}
\begin{align}
    g_{\vec r}^{(X)} &= X_{\vec r}\, 
                       X_{\vec r + \hat{\vec y}}\, 
                       X_{\vec r + \hat{\vec z}}\,
                       X_{\vec r + \hat{\vec y} + \hat{\vec z}},\\
    g_{\vec r}^{(Y)} &= Y_{\vec r}\, 
                       Y_{\vec r + \hat{\vec x}}\, 
                       Y_{\vec r + \hat{\vec z}}\, 
                       Z_{\vec r + \hat{\vec x} + \hat{\vec z}},\\
    g_{\vec r}^{(Z)} &= Z_{\vec r}\, 
                       Z_{\vec r + \hat{\vec x}}\, 
                       Z_{\vec r + \hat{\vec y}}\, 
                       Z_{\vec r + \hat{\vec x} + \hat{\vec y}}.
\end{align}
\label{eq:non_css_plaquette_checs}
\end{subequations}
The geometry of stabilizers and logical operators is also very similar to that sketched for the CSS version in \autoref{fig:plaquette_codes_geometry}. We obtain one more set of string-like Pauli-$Y$ stabilizer generators, so that now we have $3(L-1)^2$ independent string-like stabilizer generators for both both odd and even system sizes. As in the CSS case, the stabilizer group of the code has sheet-like elements for even $L$, however these sheets are not independent for different planes anymore since one can (for even $L$) use products of e.g. string-like  $X$ and $Y$ stabilizers to obtain the plane-like $Z$ stabilizers. 

In summary, the non-CSS plaquette codes are a family of codes with
\begin{equation}
    [[N, K, D]] = \begin{cases}
        [[L^3, 3L-2, L]] & \text{if $L$ even}\\
        [[L^3, 3L-1, L]] & \text{if $L$ odd}
        \end{cases}.
\end{equation}
and $3(L-1)^2$ independent stabilizer generators for odd system sizes and $3(L-1)^2+1$ independent stabilizer generators for even system sizes.

\bibliography{references.bib}

\end{document}